\documentclass[12pt,preprint]{aastex}
\usepackage[dvips]{color,epsfig,rotating}
\usepackage{graphics}

\newcommand{\bvec}[1]{\mbox{\boldmath$#1$}}

\begin{document}

\title{The Fundamental Plane of QSOs and the Relationship Between Host
and Nucleus}

\shorttitle{The QSO Fundamental Plane}

\author{Timothy S. Hamilton,\altaffilmark{1,2,3,4,5}
Stefano Casertano,\altaffilmark{3,5}
and David A. Turnshek\altaffilmark{2,5}}

\altaffiltext{1}{Research performed while a National Research Council Associate at NASA/GSFC}

\altaffiltext{2}{Dept. of Physics \& Astronomy, University of Pittsburgh,
Pittsburgh, PA 15260, USA}

\altaffiltext{3}{Space Telescope Science Institute, 3700 San Martin Drive, 
Baltimore, MD 21218, USA}

\altaffiltext{4}{Present address: Shawnee State Univ., Dept. of Natural Sciences, 940 2nd St., Portsmouth, OH 45662}

\altaffiltext{5}{email: thamilton@shawnee.edu, stefano@stsci.edu, 
turnshek@pitt.edu}

\begin{abstract}

We present results from an archival study of 70 medium-redshift QSOs
observed with the Wide Field Planetary Camera 2 on board the {\it
Hubble Space Telescope}.  The QSOs have magnitudes $M_V\leq -23$
(total nuclear plus host light) and redshifts $0.06\leq z\leq 0.46$.
The aim of the present study is to investigate the connections between the 
nuclear and host properties of QSOs, using high-resolution images 
and removing the central point source to reveal the host structure.  
We confirm that more luminous QSO nuclei are found in more luminous host
galaxies.  
Using central black hole masses from the literature, we find that nuclear luminosity 
also generally increases with black hole mass, but it is not tightly correlated.
Nuclear luminosities range from 2.3\% to 200\% of the 
Eddington limit.  Those in elliptical hosts cover the range fairly evenly, 
while those in spirals are clustered near the Eddington limit.
Using a principal components analysis, we find a kind of fundamental 
plane relating the nuclear luminosity to the size and effective 
surface magnitude of the bulge.  
Using optical nuclear luminosity, this relationship explains 96.1\% of
the variance in the overall sample, while another version of the 
relationship uses x-ray
nuclear luminosity and explains 95.2\% of the variance.  
The form of this QSO fundamental plane shows similarities to the
well-studied fundamental plane of elliptical galaxies, and we examine
the possible relationship between them as well as the difficulties
involved in establishing this connection.  

\end{abstract}


\section{INTRODUCTION}\label{sec:intro}

The availability of high-resolution, space-based imagery has been of great help to 
the study of QSOs and their host galaxies.  
Many low-to-medium redshift QSO hosts 
have been imaged with the Hubble Space Telescope in the 
past several years.  
At the same time, there has been a growing understanding of the workings 
of galactic nuclei and their central black holes.  Magorrian et al.~(1998) have found 
a proportionality between black hole and galactic spheroid masses in normal 
galaxies, and this has 
been refined by Kormendy \& Gebhardt~(2001), 
H\"aring \& Rix~(2004), and McLure et al.~(2006).  
McLure \& Dunlop~(2001, 2002), Marconi \& Hunt~(2003), and Graham~(2007) 
have studied this relation for active galaxies.  
Black hole mass has also been found to be related to the central stellar velocity 
dispersion (Ferrarese \& Merritt~2000; Gebhardt et al.~2000; 
Tremaine et al.~2002; Ferrarese \& Ford~2005; Bernardi et al.~2007).  
The velocity dispersion is itself tied to larger-scale galaxy properties 
through the fundamental plane of elliptical galaxies (Djorgovski \&
Davis~1987; Dressler et al.~1987).  So there is now a wealth of 
evidence emerging on the individual relationships between large-scale 
properties of galaxies and those of their central regions.

In this paper we report on the connections between the nuclear and
host properties of a large sample of low-redshift QSOs observed with
the Wide Field and Planetary Camera 2 (WFPC2) on board
the {\it Hubble Space Telescope} ({\it HST}).  This
follows our study of the QSO host luminosity function, described by 
Hamilton et al.~(2002).  We have collected and reanalyzed wide-band
archival images of 70 QSOs with $M_V \leq -23$ mag (total nuclear plus
host light) and redshifts $0.06 \leq z \leq 0.46$.  
This includes images taken by 
several teams, especially Bahcall et al.~(1997), Hooper et al.~(1997), 
Boyce et al.~(1998), and McLure et al.~(1999), among others.  We have taken an
inclusive approach in our sample selection, imposing no additional
selection criteria on the QSOs besides those of total absolute
magnitude and redshift.  
For each object we have subtracted the nuclear light component using
two-dimensional image fits and have derived the luminosity and size of
the underlying host galaxy by fitting an elliptical or disk-like light profile.  
A comparison of our results with other teams using the same data is given in 
\S\ref{sec:appendix-a}.

We examine the host and nuclear luminosities in \S\ref{sec:nucvshost} and confirm 
earlier results that more luminous QSO nuclei are found in more luminous host
galaxies.  The range of host magnitudes is narrower than that of the nuclei, 
and we discuss possible selection effects.  
In \S\ref{sec:nuclum}, we compare literature estimates of central black 
hole masses with our nuclear and bulge luminosities.  Nuclear luminosity 
generally increases with black hole mass, though there is not a strong correlation.  
Bulge luminosity shows little trend at all.  We find the nuclear bolometric 
luminosities span a wide range of Eddington fractions, except among QSOs in 
spiral hosts, which cluster near the Eddington limit.  

Using a principal components analysis, we 
demonstrate the existence of a fundamental plane connecting 
the nuclear luminosity to the host bulge's size and effective surface magnitude 
(\S\ref{sec:fp}).  This QSO fundamental plane can be found using either the 
optical or x-ray nuclear luminosities.  The optical version explains 96.1\% of the 
sample's variance, and the x-ray version explains 95.2\%.  When we 
look at particular subsamples of QSOs, the plane tilts in different orientations, 
which are examined in \S\ref{sec:appendix-b}.

Throughout this paper, we adopt
a Friedmann cosmology with $H_0=50$ km s$^{-1}$ Mpc$^{-1}$, $q_0=0.5$,
and $\Lambda=0$, to be in keeping with Hamilton et al.~(2002).  The effects on the results of
assuming a more recent cosmology, $H_0 = 65$ km s$^{-1}$ Mpc$^{-1}$, $\Omega_{\Lambda}=0.7$, $\Omega_{\mathrm{matter}}=0.3$,  
are discussed in \S\ref{sec:disc-fp}.

\section{DATA}\label{sec:data}

\subsection{Sample}

The sample is composed of 70 archival {\it Hubble Space Telescope}
({\it HST}) images of low-redshift QSOs.  The selection criteria are
that they have redshifts between $0.06 \leq z \leq 0.46$ and total
(host plus nucleus) absolute magnitudes brighter than $M_V \leq -23$.
Furthermore, they must have been observed with the
{\it HST}'s Wide-Field Planetary Camera 2 (WFPC2), using broad-band
filters, and have images publicly available in the {\it HST} archives
as of 1999.  This brings our sample to 70 QSOs.

Rather than restrict our study to a specific class of QSOs, we impose
no physical criteria on the QSOs beyond those of magnitude and
redshift.  Thus we are able to study a broad range of properties and
draw general conclusions.

\subsection{Image Analysis}\label{sec:imageanalysis}

Even at {\it HST} resolution, the light of the unresolved nuclear
central source significantly affects the extended light distribution
of the host galaxy.  A careful subtraction of the central point source
is needed in order to measure the properties of the host accurately.
The image reduction is described in detail by Hamilton et al.~(2002).  
The following is a brief summary of the method,
which is largely similar to that of Remy et al.~(1997).

Because of the complex structure of the {\it HST} WFPC2 Point Spread
Function (PSF), our analysis procedure has three principal steps: (1)
A two-dimensional model of the PSF is fitted to the central point
source, in order to determine its subpixel position and the telescope
focus, which affect the shape of the PSF.  (2) The PSF and a galaxy
model are simultaneously fitted to the entire image to distinguish the
nuclear and host components.  (3) The nuclear magnitude is determined
from the PSF model.  Then the fitted PSF is subtracted, and the
magnitude of the host is determined from the residual light.

The model PSF is constructed from a set of artificial PSFs, created
using the TinyTim software (the latest version is described by Krist \& Hook~2004).  
Pixel sampling is important (especially for the Wide Field images) 
because in the shape of the response function, the PSF structure
changes significantly, depending on both the telescope focus and on
exactly how the point source is centered with respect to the pixel
grid.  For instance, a point source centered in the middle of a pixel 
will look different from one centered on a corner.  If this centering is 
not accounted for, the model will not correctly apportion the light between 
the PSF and the extended source.  The WFPC2 camera used for each 
observation is listed in Hamilton et al.~(2002), Table~1.

The general PSF model is first fitted to the core of the QSO
image, where the nuclear point source is the dominant feature, with
the light of the extended galaxy treated as a constant background term.
Saturated pixels and those affected by CCD ``blooming'' are masked
from the fits.  If the PSF is not saturated, we can achieve a
precision of $\approx 0.1$ pixels in the central position and
$\approx 1 \mu\rm m$ in the focus position, for QSOs dominated by their
nuclei.  Our technique works in the presence of saturation, although
the focus is determined less accurately.  Once the position and focus
have been found, a PSF of angular size large enough to cover the host
image is created with these parameters, and it is used in the
subsequent analysis.

A second two-dimensional fit distinguishes the light of the QSO from
that of the resolved host galaxy, simultaneously fitting both parts.
In this step, the model PSF's brightness is scaled to match the QSO
nuclear brightness, while a galaxy model is fitted to the host.  The
host model includes a simple morphological classification based on
radial profile.

For spiral hosts with a visible bulge, we use an elliptical mask to fit the bulge
and disk separately.  
The mask's shape is determined by the QSO's 
isophotal ellipticity and orientation.  
It exposes the region in which the bulge dominates the disk.  
The bulge+PSF is fitted first, using the mask.  Next, the bulge model is
subtracted from the entire image, and finally the disk is fitted.  
Bars and any other irregular components are not modeled; 
they have masks drawn by eye.  This includes the irregularly-shaped bulges in 
PG~0052+251, MRK~1014, PG~1001+291, 
LBQS~1222+1010, and PG~1402+261.  
Of the regular bulges, we note that some are best modeled by a disk-like radial
profile, while others are best fitted with an elliptical galaxy type
profile.

  Based on visual inspection, some hosts appear to have undergone recent,
strong interactions that have severely distorted their appearance from
those of a normal elliptical or spiral, and we have noted these in
Table~\ref{tab:fullsample}.  The ``interacting'' 
category is rather subjective, and if seen at higher resolution, many others might 
have been classed this way.  

Using the fitted parameters, we then subtract the properly scaled PSF
from the QSO image, leaving the host galaxy.  The magnitude of the
nucleus is measured directly from the scaled PSF model.  The host
magnitude is measured from the PSF-subtracted image, within an
aperture large enough to encompass the visible extent of the host.
Outside the aperture, we extrapolate the host model to a radius of
infinity and add this contribution to the light contained within the
aperture, yielding the apparent magnitude of the host galaxy.  The
size of the host is represented by the half-light radius, $r_{1/2}$,
in the case of ellipticals or the effective radius,
$r_{\mathrm{eff}}$, in the case of spirals.  This parameter is taken
directly from the fitted host model.

The absolute host and nuclear magnitudes are calculated as described
in \S3.3 of Hamilton et al.~(2002); we briefly summarize the procedure here.  From the filter magnitude measured above, we first obtain an 
apparent $V$ magnitude by applying a color correction.  Colors
are taken from Cristiani \& Vio~(1990), for the nuclei, and from Fukugita et
al.~(1995), for the hosts.  Cristiani \& Vio~(1990) use the Johnson filter set, 
so we convert the nuclear colors to the WFPC2 filters, using the IRAF {\sc synphot} 
package and a power-law spectrum of the form $f_\nu = \mathrm{constant}$.  
Once we have the apparent $V$ magnitudes, the absolute $V$ magnitudes
are given by
\begin{equation} 
M_V = m_V - 45.396 - 5 \log (1+z - \sqrt{1+z}) - K(V) - A_V \mbox{ ,}
\end{equation}
where $K(V)$ is the $V$-band K-correction, and $A_V$ is the Galactic
extinction.  Nuclear K-corrections are also taken from Cristiani \&
Vio~(1990).  For host K-corrections, we use the data of Pence~(1976), 
which assumes no intrinsic reddening in the host galaxies.  
Galactic extinction, $A_V$, comes from Schlegel, Finkbeiner, \&
Davis~(1998), by way of the Galactic extinction calculator on
the NASA Extragalactic Database \footnote{\tt nedwww.ipac.caltech.edu}
(NED).

The error bars used in the plots and listed in Table~\ref{tab:pcasample} are 
the 1$\sigma$ statistical uncertainties.
In looking at the systematic errors, tests on simulated QSOs (including elliptical, disk, and disk+bulge hosts) 
show that we can recover their parameters rather well.  
In these tests, nuclear magnitudes are particularly accurate, with errors consistently less than 0.1 mag, 
and comparing favorably with the statistical uncertainties of $\approx 0.15$ mag.  
Host magnitude errors are $\lesssim 0.2$ mag, extending a bit larger than the statistical uncertainties, which are typically $\sim 0.1$ mag.  
Effective surface magnitudes are generally off by $\sim 0.2$ mag arcsec$^{-2}$.  The distribution of their statistical uncertainties peaks at $\approx 0.12$ 
mag arcsec$^{-2}$.  
And the errors in log radius ($\log r_{1/2}$ or $\log r_{\mathrm{eff}}$) are broadly distributed, almost all $\lesssim 0.02$, 
ranging a bit larger than the typical statistical error of $\approx 0.01$.  

The effect of PSF convolution is probably the major source of error in these tests.  
Mathematically speaking, the light entering 
the telescope is first convolved with the telescope's PSF 
and then binned into pixels.  
Our code convolves the host model with the PSF as it performs the fit, to better match 
the ideal, unconvolved parameters.  
But the effect of the finite aperture of the telescope is that some information is still lost.
Simulations can show us systematic errors, but we can also look at systematic differences 
between our analysis method and those of other groups, simply by comparing our results.  
These are discussed in \S\ref{sec:appendix-a}.

\subsection{X-ray Luminosities}

The x-ray luminosities, listed in Table~\ref{tab:pcasample} for a subset of the objects, are
obtained from {\it ROSAT} and {\it
Einstein} satellite data.  For the {\it ROSAT} observations of Brinkmann
et al.~(1997) and Yuan et al.~(1998), the data are provided as flux,
$f_x[0.1,2.4\, \mathrm{keV}]$, in units of mW m$^{-2}$ 
integrated over the energy range 0.1---2.4 keV in the observer's 
frame.  Also listed is the energy spectral index, $\Gamma$, under the 
assumption of a power-law spectrum.  With a spectrum of the form
\begin{equation}
	f_\nu = A_x \nu^{\alpha_x}
	\mbox{ ,}
\end{equation} 
where $A_x$ is a constant and $\alpha_x \equiv - (\Gamma - 1)$, 
then the constant of proportionality is
\begin{equation}
	A_x = \frac {f_x \left[ E_{\mathrm{low}}, 
	E_{\mathrm{high}} \right] \, (1+\alpha_x)} 
	{\nu^{1+\alpha_x}_{\mathrm{high}}
	- \nu^{1+\alpha_x}_{\mathrm{low}}} 
	\mbox{ ,}
\end{equation}
where $E_{\mathrm{low}} = 0.1$ keV, and $E_{\mathrm{high}} = 2.4$ keV.
The frequencies correspond to the energies as $\nu_{\mathrm{high}} =
1000 E_{\mathrm{high}}/h$ (similarly for $\nu_{\mathrm{low}}$), 
where $h$ is Planck's constant.  The luminosities, $\nu L_\nu
\mid_{\nu_0}$, are evaluated at a rest-frame energy of $E_0 = 0.5$ keV
(with $\nu_0 = 1000 E_0/h$).
The x-ray luminosity for PG~1229+204 comes from the PSPC-pointed {\it ROSAT} observation 
reported by Grupe et al.~(2001).  Reduction is performed as above, 
but over the 0.2---2.0 keV range.

For x-ray data from the {\it Einstein} satellite Imaging Proportional
Counter (Wilkes et al.~1994; Margon et al.~1985), the energy range used is
0.16---3.5 keV.  The luminosity of 3C~93 is calculated 
from the flux listed by Wilkes et al.~(1994) for a 
spectral index of $\alpha_x \approx -1.5$, which is the closest 
to the average of our sample.  
Margon et al.~(1985)
actually provide the Q~2344$+$184 x-ray luminosity itself, in the
format desired here, but they do not quote a spectral index.  
Regardless of the source of data, we use the notation $L_X$
for the nuclear x-ray luminosities in ergs s$^{-1}$,
evaluated at the rest frame energy of 0.5 keV.  

\subsection{QSO Classifications}

Radio-loudness classifications are collected primarily from Brinkmann et
al.~(1997) and Yuan et al.~(1998), both of whom use a loudness
criterion that classifies an object as radio-loud if it has a
radio-to-optical flux density ratio in excess of 10.  Radio-loudness
classifications for the remaining objects come from a variety of sources, with
extensive use made of the NASA Extragalactic Database.

In the analyses that follow, the QSO sample is subdivided into several
categories, on the basis of radio-loudness and host morphology.  We
refer to radio-loud (L) and radio-quiet (Q) QSOs, as well as those
with elliptical (E) or spiral (S) hosts.  These categories are
combined to describe radio-loud QSOs in elliptical hosts (LE),
radio-louds in spiral hosts (LS), radio-quiets in elliptical hosts
(QE) and radio-quiets in spiral hosts (QS).  We note that the LS
subsample has just four members, so it is omitted from many of the
statistical analyses that follow.  The more detailed 
morphological descriptions applied to some hosts in Table~\ref{tab:fullsample} 
(such as ``interacting'') are not used in the analyses; only the 
E and S descriptions matter.

\subsection{Redshift Effects}

We find some redshift effects on the measured parameters.  
QSOs with nuclei much fainter than the hosts (up to 3.5 mag fainter) are only seen at low redshifts ($z < 0.3$).  
At high redshifts ($z \geq 0.3$), the nuclear absolute magnitudes do not get more than 
about 0.5 mag fainter than the host.  
The median effective surface magnitude ($\mu_e$, described in detail in \S\ref{sec:surfmagcalc}) 
brightens from 21.48 to 20.86 mag arcsec$^{-2}$ as we go to $z \geq 0.3$, 
and we see somewhat fewer faint hosts at high redshift, though the trend is not strong.
We do not, however, see a trend in $\log r_{1/2}$ itself (when measured in kpc).  

The uncertainties in host and nuclear magnitudes show no change with redshift.
The uncertainties in $\log r_{1/2}$ and $\mu_e$ are larger, on average, 
for objects with redshift above 0.3.  The median $\sigma_{\log r_{1/2}}$ rises from 0.007 (for $z<0.3$) to 0.023 (for $z \geq 0.3$), 
while the median $\sigma_{\mu_e}$ rises from 0.10 to 0.14.

\section{NUCLEAR VS.\ HOST LUMINOSITIES}\label{sec:nucvshost}

\subsection{Expectations}

In addition to quantifying the behavior of nuclei and hosts
separately, a major purpose for decomposing the QSO image into nuclear
and host components is to look for relationships between the two.  The
correlations between nuclear and host parameters are important because
they shed light on the interplay between the central engine and the
galaxy that harbors it.  One of the most straightforward questions is
whether or not the nuclear and host luminosities are correlated.

In posing the question, we will assume the standard model of the
central engine, a supermassive black hole surrounded by an accretion
disk.  Further assuming a fixed mass-to-light ratio for galaxies 
(following, for example, J{\o}rgensen et al.~1996 and Magorrian et
al.~1998), then the more massive a galaxy is, the more luminous it is.
The mass of the galaxy includes not only stars but also gas, in the
form of nebulae and a diffuse interstellar medium.  Even in elliptical
galaxies, which have a lower gas content than spirals, there can be
gas in the form of a hot interstellar medium given off by the winds of
post-main-sequence stars.  Some of this gas (of both types) eventually
accretes onto the central engine as fuel (Di Matteo et al.~2000), so a
more luminous galaxy has more fuel available for the nucleus.  Under
Bondi accretion (Bondi~1952), the gas accretion rate onto the disk is
proportional to the density of the gas far from the engine and to the
square of the black hole mass, and the luminosity of the accretion
disk is proportional to the accretion rate (Carter~1979).  So given
the model assumed here, a correlation between nuclear and host
luminosities may naively be expected.

Earlier investigations have shown evidence
of such a relationship.  Hutchings et al.~(1984)
studied 78 QSOs and lower-luminosity Active Galactic Nuclei (AGN) out to a redshift of $z=0.7$
and found a positive correlation between these parameters, suggesting
that nuclear power is related to galaxy mass.  McLeod \& Rieke~(1994a,
b) used infrared $H$-band imaging of QSOs, in which band the ratio of
nuclear to host luminosity is diminished.  They considered two samples
of low-redshift QSOs, one with low total (nucleus plus host)
luminosities and one with high total luminosities, and they found that
the hosts of the high-luminosity sample are brighter than those of the
low-luminosity sample.

\subsection{Analysis}

The extinction-corrected absolute nuclear and host magnitudes are
plotted against each other in Figure~\ref{fig:mnuc-mhost1}, with
separate symbols for the LE, QE, LS, and QS categories.  
The object MS~2159.5$-$5713 is plotted as a separate symbol, because we do not know its
radio-loudness.

The dotted curve in Figure~\ref{fig:mnuc-mhost1} represents
approximately the location of objects with a combined nuclear plus
host magnitude of $M_V\mathrm{(total)} = -23.0$, the faintest allowed
under the conventional definition of a QSO and therefore the faint limit for
our sample.  The sample selection is made using published,
ground-based data, so this limit is only approximate.  The region in which 
the nuclear absolute magnitude is
$M_V\mathrm{(nuc)}>-23$ suffers from selection effects.  We
therefore perform the fits and correlation tests in this section only 
on those QSOs for which $M_V\mathrm{(nuc)} \leq -23$.

\subsubsection{Correlations and Functional Forms}

The Spearman correlation coefficients for $M_V\mathrm{(nuc)}$ and
$M_V\mathrm{(host)}$ are listed in Table~\ref{tab:spear-fits}.
The overall correlation is not strong ($\rho=0.350$), but we see greater
correlations among radio-loud QSOs ($\rho=0.578$) and especially among
the 4 members of the LS subsample ($\rho=0.800$).  Unfortunately, the
small number of LS QSOs makes it difficult to interpret this result.

We can also describe the data by fitting functional forms (Table~\ref{tab:spear-fits}).  
For this we use the ``Bivariate
Errors and intrinsic Scatter'' (BES) estimator, described by Akritas
\& Bershady~(1996).  The BES method accounts for errors in both
variables and is therefore useful in this application, for which there
is not a perfectly-measured independent variable.  The parameter
$M_V\mathrm{(nuc)}$ is used as the independent variable because it has
a larger range than $M_V\mathrm{(host)}$ does, even after restricting
it to $M_V\mathrm{(nuc)} \leq -23$.

The functional fit to the overall sample, shown as the solid line in
Figure~\ref{fig:mnuc-mhost1}, can be expressed as
\begin{equation}
	M_V\mathrm{(host)} = (0.303 \pm 0.088)
	M_V\mathrm{(nuc)}-(15.9 \pm 2.2) \mbox{ .}
\end{equation}
That is, host luminosity varies with nuclear luminosity as a power law
with an exponent of 0.3.  The slopes of most of the subsamples are
similar to this, to within the errors.  Those subsamples with
especially low $M_V\mathrm{(host)}$ to $M_V\mathrm{(nuc)}$
correlations, QS and Q, are exceptions, as is the unreliable LS
subsample.

\subsection{Discussion}

It appears from this analysis that QSOs with more luminous nuclei are
slightly more likely to have more luminous hosts.  The correlation
itself is rather weak ($\rho=0.350$) but unlikely to have occurred by
chance (probability $\approx 0.008$).  The luminosity range we consider,
$M_V\mathrm{(nuc)} \leq -23$, is chosen to eliminate the stronger
selection effects.

\subsubsection{Selection Effects}

The extremely narrow range of $M_V\mathrm{(host)}$ as compared to
$M_V\mathrm{(nuc)}$ is partly due to the sample selection, with
objects being selected to have a combined $M_V\mathrm{(nuc+host)} \-
\leq -23$.  But even when just
the $M_V\mathrm{(nuc)} \leq -23$ region is examined, the range of
magnitudes spanned by the hosts (3.1 mag) is smaller than the range
spanned by the nuclei (4.2 mag).

If we instead used all of the objects lying outside the $M_V\mathrm{(nuc+host)} \leq -23$ line (rather than $M_V\mathrm{(nuc)} \leq -23$), our $M_V\mathrm{(nuc)}$  vs. $M_V\mathrm{(host)}$ fit would be slightly flatter (slope=0.24), but within 1-$\sigma$ of our results here.  The subsample slopes would show a similar change, except for the radio-louds, which aren't affected (they only exist at higher nuclear luminosities).
The Spearman correlation would be stronger (from $\rho=0.350$ to 0.451).  The radio-loud subsamples, again, would not be affected.  But all of the others would be strengthened, except for the QE's, which weaken (from $\rho=0.394$ to 0.191).
The correlations might be even stronger if fainter classes of AGN (those with $M_V\mathrm{(nuc+host)} > -23$) were included, but they are outside the scope of this study.

\begin{figure}
\plotone{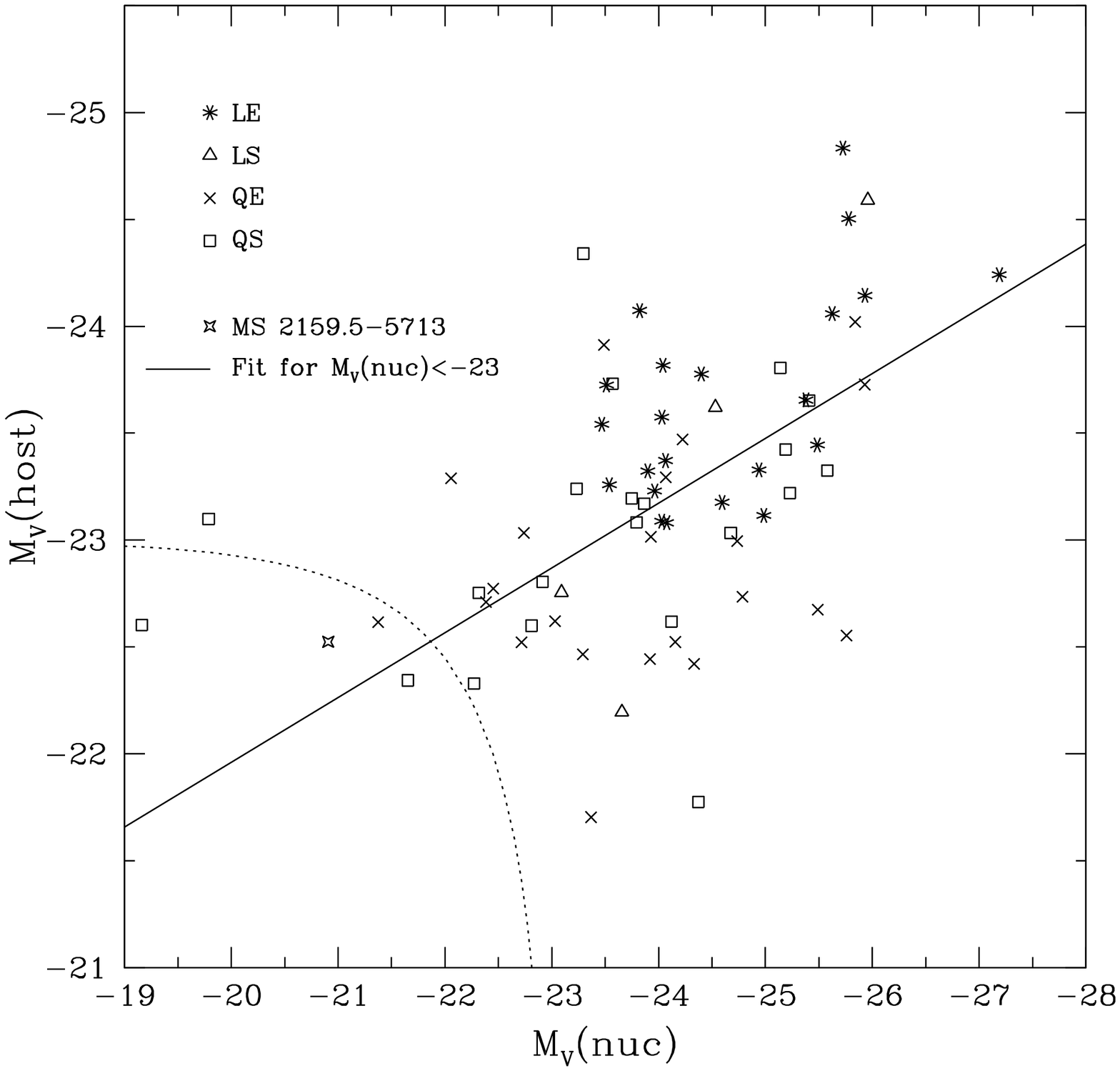}
\caption{Overall distribution of host and nuclear luminosities.
The QSOs are marked as radio-loud (L) or quiet (Q) and with elliptical (E) or spiral (S) hosts.
The dotted curve is the approximate bound for objects with total (nucleus plus host) magnitude of 
$M_V\mathrm{(total)} > -23$.  
The solid line is fitted to those QSOs with nuclear luminosity 
$M_V\mathrm{(nuc)} \leq -23$ to avoid any bias by the few objects with fainter nuclei.
}
\label{fig:mnuc-mhost1}
\end{figure}

There may be objects missing from the region of
Figure~\ref{fig:mnuc-mhost1} with high host luminosity and low nuclear
luminosity (the upper left corner), due to sample selection effects.
Optical searches for QSOs have typically stipulated that objects have
a ``stellar appearance'' on photographic plates.  The dominance of the
host here might keep this an empty region in optically-selected QSO
samples, but it is not a problem for x-ray and radio-selected samples.

This study uses archival data from a variety of {\it HST} proposals
and, therefore, a mix of QSO selection methods.  These are noted in
Table~\ref{tab:fullsample}.  From the proposals and published articles
by the original proposers, we have determined that only one object,
PG~1229+204, was chosen for its lack of an extended host in
ground-based images.  An additional 13 QSOs were chosen from
optically-selected catalogs, most commonly the Palomar-Green (PG)
survey (Schmidt \& Green~1983).  Another 9 QSOs were chosen from the
Large Bright Quasar Survey (LBQS), which is optically selected but has
safeguards to prevent the rejection of objects with detectable
``fuzz'' surrounding the nucleus (Foltz et al.~1987).  The remaining
47 QSOs in the sample were originally observed with {\it HST} for
reasons that do not have direct optical biases.  This includes radio-
and x-ray-selected QSOs.  For those QSOs that appear in multiple {\it
HST} proposals that were available for this study, we place them in
the most ``unbiased'' category their proposals permit.  The QSOs
proposed for or selected in ways most likely to be biased against the
upper left corner of Figure~\ref{fig:mnuc-mhost1} therefore number 14
out of 70 (those marked with ``B'' or ``C'' in
Table~\ref{tab:fullsample}).  Most of these 14 QSOs have nuclei that
are more luminous than their host galaxies and tend to lie on the
right-hand side of Figure~\ref{fig:mnuc-mhost1}, as expected.  
(And so there is little overlap with those 14 QSOs that are fainter than the 
$M_V\mathrm{(nuc)} \leq -23$ cut.)  But two 
of them have approximately equal host and nuclear luminosities
(PG~1229+204 and PHL~1093), and US~3498 has a nucleus 3.3 magnitudes
less luminous than its host.  

As for the effects of these potentially biasing QSOs on the
$M_V\mathrm{(nuc)}$--$M_V\mathrm{(host)}$ correlation, we note that
the farthest outliers in the upper-right corner of
Figure~\ref{fig:mnuc-mhost1} are not among the 14 biased QSOs.
Furthermore, the sample of radio-loud QSOs includes only 2 such biased
objects, and the correlation between $M_V\mathrm{(nuc)}$ and
$M_V\mathrm{(host)}$ is actually stronger for that sample.

\subsubsection{Technical Comparison with Other Studies}

Other space-based studies have examined the relationship between host and nuclear
luminosity, as well.  We present the technical details of their
observations and analysis methods here, but we will defer a comparison
with their results until \S\ref{sec:finaldisc-hostnuclum}.

Bahcall et al.~(1997) use the WFPC2 on {\it HST} to study a
sample of 20 QSOs with redshifts $z \leq 0.3$ and total absolute
magnitudes in the range $-24.4 \leq M_V \leq -27.1$.  All of these QSOs are
included in our sample.  Observations of four different
stars are used for PSFs.  The nucleus is subtracted from each image
using both 1- and 2-dimensional methods, with the 2-D results being
adopted in the end.  In the 2-D method, the PSF is subtracted from
the QSO first, with each of the four stars being tried in turn, and
the one which leaves the fewest residuals being adopted.  A host model
is then fitted to the residual image.  

Dunlop et al.~(2003) also use the WFPC2 on {\it HST} to study a sample
of 10 radio-loud QSOs and 13 radio-quiet QSOs with redshifts $0.1 \leq
z \leq 0.25$.  All of their QSOs are included in our sample except for 5 objects 
(3 radio-louds and 2 radio-quiets) whose images were not 
publicly available when we defined our sample and began analysis.  Stellar observations are
used to create a PSF that is subtracted from the QSO images using
2-dimensional modeling, with the host model being fitted
simultaneously with the scaled PSF.

Another study using {\it HST}'s WFPC2 is that of O'Dowd et al.~(2002), 
who analyze the host and nuclear luminosities of 40 BL
Lacs and 22 radio-loud QSOs with redshifts between 
$0.015 \lesssim z \lesssim 0.5$.  They perform a simultaneous host
plus PSF fit to the radial profiles of the BL Lacs.  For the QSOs,
they use the published results from other studies and therefore mostly
overlap our sample.

\section{NUCLEAR LUMINOSITY AND ITS RELATION TO BLACK HOLE 
MASS}\label{sec:nuclum}

\subsection{Black Hole Mass}\label{sec:bhmass}

Black hole masses have been variously estimated using H$\beta$
line widths, galaxy bulge masses, and stellar velocity dispersions,
among other methods, since they cannot be measured directly.  There
are problems with applying some of these methods to QSOs, given their
great distances and the glare from the nuclear emission.  
Not all are convinced by such methods, and it is with these caveats that 
we adopt the following black hole masses from the literature.  
But while the mass measurements themselves are subject to these systematic uncertainties, 
their use by other researchers does permit us to make self-consistent comparisons 
of their QSO results with ours.  

Magorrian et al.~(1998) study black holes in nearby, normal galaxies.
Using {\it HST} photometry and ground-based spectroscopy, they model
the stellar orbital velocities for given mass-to-light ratios and
black hole masses.  With the assumption that the black hole mass is proportional to bulge mass, 
they find that $\mathcal{M}_{\mathrm{BH}} = 0.005 \mathcal{M}_{\mathrm{bulge}}$, 
where $\mathcal{M}_{\mathrm{BH}}$ represents the black hole mass and
$\mathcal{M}_{\mathrm{bulge}}$ is the bulge mass.  
More recently, other groups 
(Kormendy \& Gebhardt~2001; Marconi \& Hunt~2003; McLure et al.~2006) 
have used improved measurements to revise the black hole to bulge mass ratio downward to 
\begin{equation}\label{equ:magorrian-revised}
	\mathcal{M}_{\mathrm{BH}} \sim 0.002 \mathcal{M}_{\mathrm{bulge}}
	\mbox{ .}
\end{equation}
H\"aring \& Rix~(2004) find evidence of a non-linear relationship, but 
their average ratio agrees with this value.  

Without assuming that this sort of relation holds true for QSOs, we have
taken the black hole masses for 26 of our QSOs from McLure \&
Dunlop~(2001) and list these in Table~\ref{tab:bh}.  They calculate the masses from the QSOs' H$\beta$
line width, a method that assumes the line width comes from the
Doppler-broadened motion of clouds around the black
hole.  This method requires knowing the size of the Broad-Line Region
(BLR), which they calculate from the relation of
Kaspi et al.~(2000).

The results of McLure \& Dunlop~(2001) come from a combination of
published data and new spectral observations.  We do not undertake
independent calculations of the black hole masses from the variety of
published line widths and BLR sizes because, as stressed by
Laor~(1998), published values of H$\beta$ for the same QSO often vary
greatly from one study to another.  McLure \& Dunlop~(2001) have
original H$\beta$ observations for nearly half of their sample, make
their calculations in a consistent manner, and demonstrate care where
they add literature data to their own observations, making sure that
these aren't likely to introduce large systematic errors.  We use their
black hole masses for those reasons and because of the large overlap
(26 objects) between their sample and the QSOs in our study.
As they list no uncertainties for the 
black hole masses, we adopt 
the high-end estimate of Vestergaard~(2004) for this method, 0.4 dex.  
This 1 $\sigma$ uncertainty  is larger than what we can obtain by propagating the 
known uncertainties.

\subsection{Eddington Limit}\label{sec:eddlimit}

We would also like to know if these QSOs are
radiating at close to their theoretical limits.  The Eddington limit
for the bolometric emission from the nucleus is given by (McCray~1979)
\begin{equation}
  L_{\mathrm{Edd}} \simeq 1.3 \times 10^{38} \,\,
  \frac{\mathcal{M}_{\mathrm{BH}}}{\mathcal{M}_{\sun}} \,\, 
  \mbox{erg sec$^{-1}$ .}
\end{equation}
We calculate the bolometric luminosities from
$M_V\mathrm{(nuc)}$ (listed in Table~\ref{tab:fullsample}), using the
data from Elvis et al.~(1994), who provide the ratios between
different monochromatic luminosities, $\nu L_\nu$, and the bolometric
luminosity, $L_{\mathrm{bol}}$ (their Table~17).  For the $V$-band,
\begin{equation}
	\left< \frac{\nu L_\nu}{L_{\mathrm{bol}}} \right> 
	= 14.2 \pm 5.1 \mbox{ ,}
\end{equation} 
where $\nu L_\nu$ is evaluated at $\nu=5.45 \times
10^{14}$~Hz, corresponding to a wavelength of $\lambda=5500$~\AA, the
central wavelength of the $V$ filter.

We obtain $L_\nu$ from our calculated $M_V\mathrm{(nuc)}$, which are
Vega-system magnitudes, using the zeropoints and photometric conversions 
of Voit~(1997).  In terms of $M_V\mathrm{(nuc)}$,
\begin{equation}
	\log f_\nu = -0.4 \left[ M_V\mathrm{(nuc)} + 48.62 \right] \mbox{ ,}
\end{equation}
where $f_\nu$ is the $V$-band flux density per unit frequency, in units of 
erg cm$^{-2}$ s$^{-1}$ Hz$^{-1}$.

The $V$-band luminosity density, $L_\nu$, is then calculated from $f_\nu$ and
the QSO's redshift, $z$.  Since $M_V\mathrm{(nuc)}$ is already
K-corrected, then no further cosmological correction needs to be
applied, and the luminosity density is obtained from the simple geometric relation,
\begin{equation}
	L_\nu=4 \pi d^2 f_\nu \mbox{ ,}
\end{equation} 
where $d$ is the distance.  The bolometric luminosity can then be calculated from
Table~17 of Elvis et al.~(1994).  
The nuclear luminosity and the black hole mass can be used together to
determine the Eddington fraction, $L_{\mathrm{bol}} /
L_{\mathrm{Edd}}$, listed logarithmically in Table~\ref{tab:bh}.  
Since the errors in nuclear magnitude are much smaller than those assumed 
for the black hole masses, the propagated errors in 
$\log \left( L_{\mathrm{bol}} / L_{\mathrm{Edd}} \right)$ 
all come out to be 0.4.  
Beaming of the nuclear emission would create further uncertainties.  We discuss this effect later.

As our calculated Eddington fractions depend on the black hole mass estimates of 
\S\ref{sec:bhmass}, the same caveats apply here.   
There could also be some problems in using the luminosity ratios of Elvis et al.~(1994).  Energies
above 10 keV are not included in the bolometric luminosity, due to the
unknown spectrum in that region at the time of the paper.  More
important is the estimate of the ``Big Blue Bump'' in the UV-optical
region.  As pointed out by Mushotzky~(1997), later X-ray studies
(Laor et al.~1997; Zheng et al.~1997) have indicated that the
continuum flux in this region is lower than some earlier estimates and
parameterizations (Mathews \& Ferland~1987).  However, Elvis et al.~(1994) are
not cited in this discussion, and though their mean spectrum has a gap
between approximately 1000 \AA \, and 0.1 keV, they may avoid this
problem because they do have data for the optical region and therefore
do not have to parameterize there.

\subsection{Results}

\subsubsection{Bulges and Black Holes}\label{sec:bulgeblackhole}

\begin{figure}
\plotone{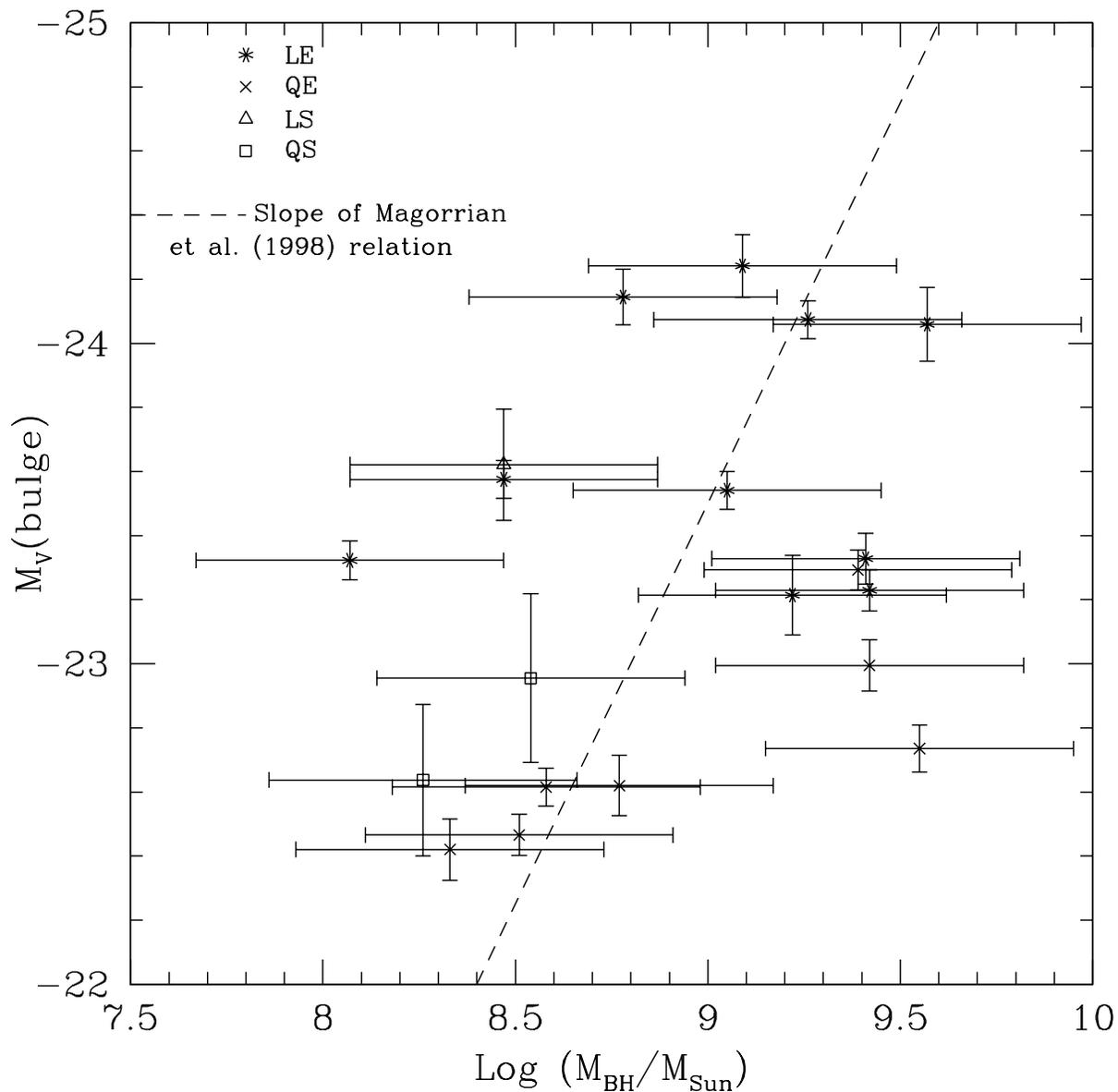}
\caption{Bulge absolute magnitude vs.\ black hole mass.  Objects with
disk-like or unmodeled bulges are not included in the plot.  
The dashed line is not a fit but is drawn parallel to the relation of Magorrian et al.~(1998). 
The correlation is weak and, given the large errors in the masses, does not resolve the question of a black hole to bulge mass relation.
}
\label{fig:mbulge-mbh}
\end{figure}

Out of the 26 QSOs that have black hole masses given
by McLure \& Dunlop~(2001), we have modeled, spheroidal bulges for 20,
whose absolute magnitudes are given in Table~\ref{tab:bh}.  
We restrict consideration to the spheroidal (elliptical) bulges, as
opposed to those that have disk-like profiles, because they are more
likely to have a dynamical relationship between their bulge and black
hole (Kormendy~2001).  In Figure~\ref{fig:mbulge-mbh}, the bulge magnitudes, 
$M_V\mathrm{(bulge)}$, are plotted against the black hole masses.
Figure~\ref{fig:mbulge-mbh} also shows the slope of 
any relationship like equation~(\ref{equ:magorrian-revised}).  
If the QSO black holes have
masses that are a constant fraction of the bulge mass, the
distribution will lie parallel to this line.  Assuming a constant
mass--to--light ratio, then if black hole mass is a fixed fraction of
bulge mass, an increase of one decade in black hole mass will
correspond to an increase in one decade in bulge luminosity, or 2.5
magnitudes.

There is a great deal of scatter in the distribution of 
bulge magnitude with black hole mass.  The two parameters show only a weak anticorrelation 
overall (Table~\ref{tab:spear-fits}), just $\rho = -0.263$.
Radio-quiet objects, on the other hand, show a stronger
anticorrelation, $\rho=-0.594$, and in particular, the 7 QE objects
have a tight anticorrelation at $\rho=-0.847$, which can be seen 
from Figure~\ref{fig:mbulge-mbh}.

The BES fits to the data are listed in
Table~\ref{tab:spear-fits}.  The overall fit can be expressed in the form
\begin{equation}
  M_V\mathrm{(bulge)} = (-1.43 \pm 1.18) \log \left( \mathcal{M}_{\mathrm{BH}} / \mathcal{M}_{\sun} \right) -
  (10.5 \pm 10.5) \mbox{ .}
\label{equ:mbulge-mbh-orig}
\end{equation} 
The slopes of most of the subsamples are similar, to within
the errors, with the notable exceptions of the spiral and QS categories.
This is not really parallel to equation~(\ref{equ:magorrian-revised}), which would have a slope of 2.5 in this figure.  The scatter is quite large, and they are about one standard deviation different.

\subsubsection{Nuclear Luminosities and Eddington Fractions}

\begin{figure}
\plotone{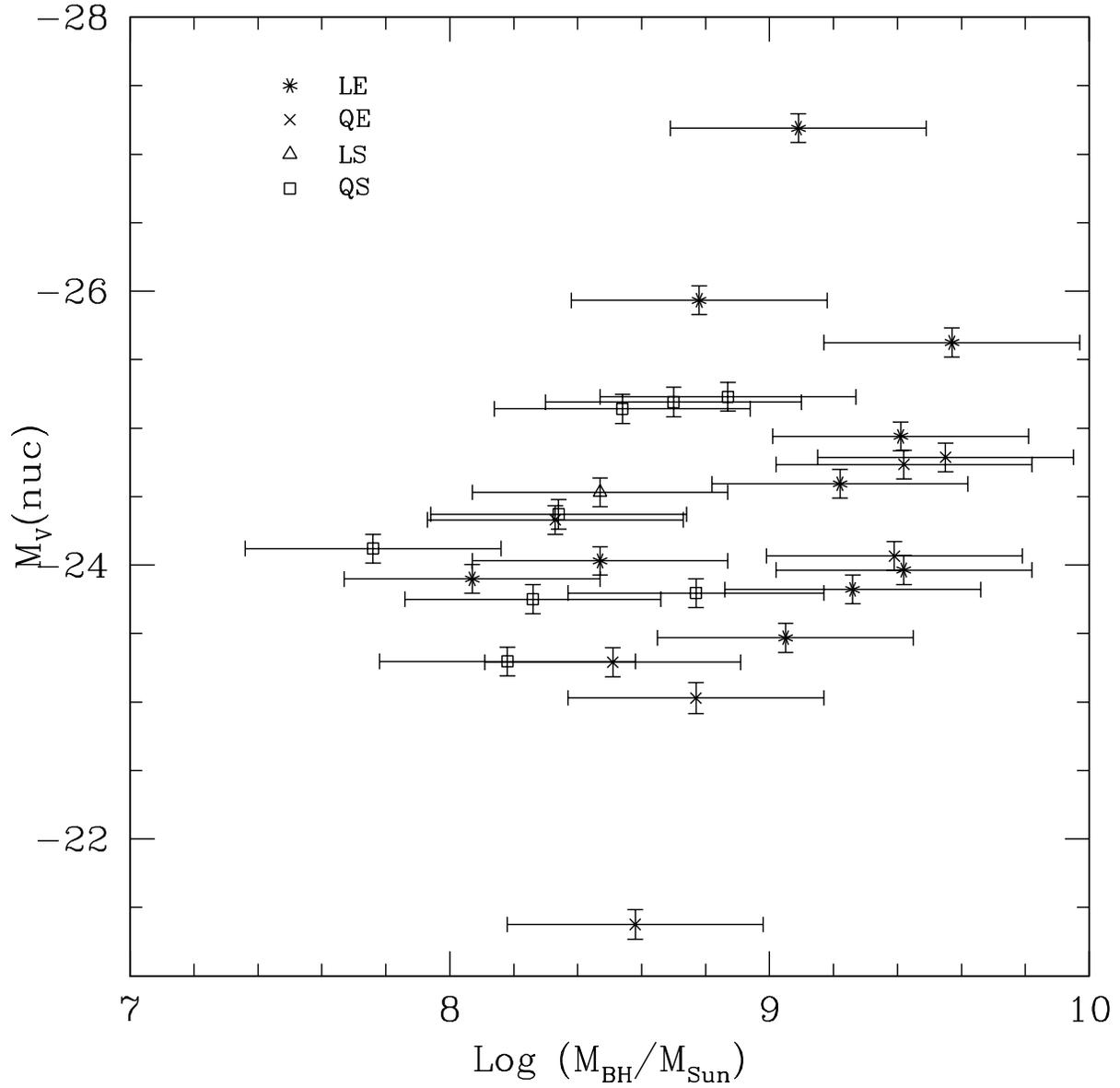}
\caption{Nuclear absolute magnitude vs.\ black hole mass.  Nuclear
luminosity generally increases with greater black hole mass, though
the correlation is low.  The lack of QSOs in the upper left
corner is due to the Eddington limit, and there might be selection
effects responsible for the lack of objects in the lower right corner.
}
\label{fig:mnuc-mbh}
\end{figure}

A plot of the nuclear luminosity vs.\ black hole mass is presented in
Figure~\ref{fig:mnuc-mbh}.  The distribution of $M_V\mathrm{(nuc)}
\mbox{ vs.\ } \log \mathcal{M}_{\mathrm{BH}}$ shows that there is a
very general trend of luminosity increasing with mass.  The Spearman
correlations, given in Table~\ref{tab:spear-fits}, include a
fairly weak anticorrelation overall, with $\rho=-0.378$.  The spiral
subsample in this case shows the strongest anticorrelation, with
$\rho=-0.667$.  Note that we have 26 objects to study in this analysis, 
because we do not impose any restrictions on the bulges, as we did in 
\S\ref{sec:bulgeblackhole}.

The fits are listed in Table~\ref{tab:spear-fits}.  The
overall fit to the data is given by
\begin{equation}
	M_V\mathrm{(nuc)} = -(1.98 \pm 1.16) \log \left( \mathcal{M}_{\mathrm{BH}} / \mathcal{M}_{\sun} \right)
			- (6.90 \pm 10.2) \mbox{ ,}
\label{equ:mnuc-mbh}
\end{equation} 
but it suffers from outlying points.  The spiral and QS subsamples show positive 
slopes, with the spirals having $2.09 \pm 2.19$.

\begin{figure}
\plotone{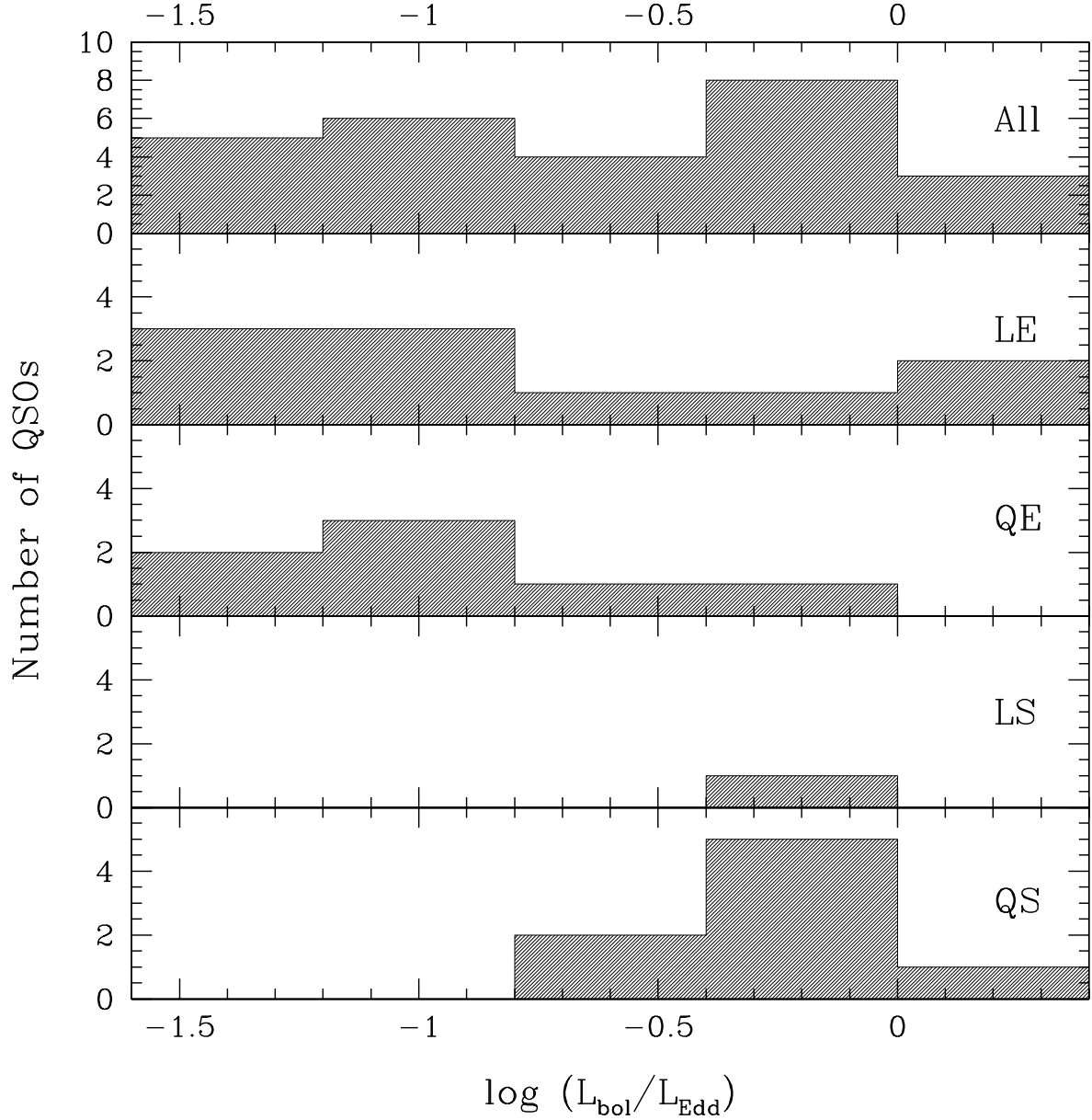}
\caption{Distributions of Eddington fractions, with a bin width of 0.4 (equal to the uncertainties) and 
subdivided by morphology and radio loudness.  
The spirals tend to cluster at high Eddington fractions, while the ellipticals are spread over a wide range.  Note the different vertical scale used for the complete sample's plot.}
\label{fig:ledd}
\end{figure}

The distribution of log Eddington fractions is presented in Figure~\ref{fig:ledd}, 
both for the complete sample and broken down by subclass.  The bin width is 0.4, 
which is equal to the uncertainties.  
In general, the QSOs demonstrate a wide range in their Eddington fractions, extending 
from $-1.6 \leq ( \log L_{\mathrm{bol}} / L_{\mathrm{Edd}}) \leq 0.3$.  
Two objects (3C~273 and PG~1402$+$261) appear to have 
luminosities above the Eddington limit, but they exceed it by less than the uncertainties.  
PG~1302$-$102, with $\log L_{\mathrm{bol}} / L_{\mathrm{Edd}} = 0.0$, is also placed in the highest bin.

We can see that the objects in spiral hosts tend to have 
higher Eddington fractions, -0.8 to 0.3 (in the logarithm), while those in elliptical hosts are 
spread across -1.6 to 0.2 (again in the logarithm).
No distinction is apparent between the Eddington fractions of radio-loud 
and radio-quiet objects.

\subsection{Discussion}

\subsubsection{Bulges and Black Holes}

We cannot say much about a fixed black hole to bulge mass ratio, 
such as equation~(\ref{equ:magorrian-revised}).  
While it appears from Figure~\ref{fig:mbulge-mbh} that there might be some weak 
correlation between the bulge magnitude and the black hole mass, 
the details of the relationship are hidden by the large uncertainties in the masses.  
The range of masses is not much larger than their uncertainties, 
allowing for a wide range of possible slopes, so the results are simply inconclusive.  
That the 7 QE objects
form the tightest correlation is interesting.  Since there are not
many objects in this category, not too much can be drawn from this
result yet, but it presents an avenue for follow-up studies.

\subsubsection{Nuclear Luminosities and Eddington Fractions}\label{sec:nuclum-eddfrac}

We see that as black hole mass increases, the QSO's nuclear luminosity
also increases (Figure~\ref{fig:mnuc-mbh}), although the trend is rather weak.  
A stronger trend might naively be expected, as the more massive black holes 
have higher Eddington limits.  Yet the QSOs do not all radiate near their limits (Figure~\ref{fig:ledd}), 
although there is a clear division according to host morphology.  Those in spiral hosts 
do tend to cluster at relatively high Eddington fractions and only go as low as 16\% of Eddington.  
But those in elliptical hosts are more evenly distributed over a wider range, extending as low as 2.5\% of Eddington.  
The difference may be caused by spirals typically having smaller black holes.

\section{THE ``FUNDAMENTAL PLANE'' OF QSOS}\label{sec:fp}

Instead of only looking at correlations of two parameters at a time, 
we can analyze multiple dimensions at once.  Principal Components 
Analysis (PCA) provides a convenient method for 
identifying and examining multidimensional correlations.  
We use this method to search for further connections between 
the nuclear and host properties.

\subsection{Analysis and Results}

\subsubsection{Restricted Sample}

For our PCA, we use a restricted sample of those QSOs for which we
have all of the following parameters: $M_V\mathrm{(nuc)}$,  $L_X$, 
$r_{1/2}$, and $\mu_e$, where $\mu_e$ is the effective
surface magnitude of the galactic bulge.  We further
require that each QSO have a modeled, spheroidal bulge (the entire
galaxy, in the case of elliptical hosts).  These qualifications
restrict the sample to the 42 QSOs listed in Table~\ref{tab:pcasample}.

In order to prevent the choice of units
from artificially weighting some parameters more than others, each
parameter is first normalized by subtracting its mean and dividing by
its variance.  The PCA is then performed on these normalized
variables.  Note that the normalization for each of the subsamples is
carried out separately from the others.  In the output, each of the
eigenvectors is written as a linear combination of the original (but
normalized) parameters.  The eigenvalues are scaled so that the sum of
all eigenvalues equals the total number of eigenvectors (and therefore
the total number of parameters as well).

\subsubsection{Effective Surface Magnitude}\label{sec:surfmagcalc}

There is already a well-studied
fundamental plane (FP) for normal, elliptical galaxies (Djorgovski \&
Davis~1987; Dressler et al.~1987) that incorporates galaxy size, $r_{1/2}$, central velocity dispersion,
$\sigma_c$, and effective surface magnitude,
$\mu_e$.  Here we define $\mu_e$ as the surface magnitude (expressed
in magnitudes arcsec$^{-2}$) of an elliptical galaxy or bulge at a distance $r_{1/2}$ from
the center of the galaxy.  Note that there is another form of effective surface
magnitude, $\left< \mu \right>_e$, which is the average surface magnitude
within the half-light radius.  These two forms are related such that $\mu_e - \left< \mu
\right>_e = 1.39$ (van Albada et al.~1993).  
We can derive an expression for
$\mu_e$ by first noting that the total host luminosity in
counts, $L_{\mathrm{host}}$ (corresponding to the absolute magnitude, 
$M_V(\mathrm{host})$), of a spheroidal galaxy following an $r^{1/4}$
profile and extended out to an infinite radius can be expressed as
\begin{equation}
	L_{\mathrm{host}} = \kappa_L \, \mathrm{e}^{7.67} \,\, I(r_{1/2})
	    \,\, r_{1/2}^2 \,\, (1-\epsilon)\mbox{ ,}
\end{equation}
where $I(r_{1/2})$ is the galaxy's surface brightness (in count-rate per unit
area) at the half-light radius, $r_{1/2}$, and
$\epsilon$ is the ellipticity of the isophotes.  Then $\kappa_L$ is a
constant, which we calculate numerically as
$\kappa_L=0.01054$.  Note that Malkan~(1984) gives an expression for
$L_{\mathrm{host}}$ that leads to $\kappa_L=0.01058$, a difference of only
0.4\%.

Converting to magnitudes arcsec$^{-2}$, we obtain
\begin{equation}
	\mu_e = M_V\mathrm{(host)} + 2.5 \log \left
                           [ \theta^2_{1/2} \, \kappa_L \,
                           \mathrm{e}^{7.67} \, (1-\epsilon) \right]
                           \mbox{ ,}
\label{equ:surfmag}
\end{equation}
where $\theta_{1/2}$ is the half-light radius expressed in arc seconds.  
In the case of an elliptical bulge, we substitute $M_V\mathrm{(bulge)}$ in place of $M_V\mathrm{(host)}$.  We can now look for a QSO fundamental plane that involves the host parameters, $\log r_{1/2}$ and $\mu_e$, as well as a measure of the nuclear luminosity, $M_V\mathrm{(nuc)}$ or $\log L_X$.
In this, we use the effective surface magnitude, rather than the bulge magnitude, so that we can make direct comparisons between our results and the well-studied fundamental plane of elliptical galaxies.

\subsubsection{PCA Results and Derivation of the QSO Fundamental Plane}

We can perform two PCAs, an optical one using $M_V\mathrm{(nuc)}$,  
$\log r_{1/2}$, and $\mu_e$ as the parameters, and an x-ray one that 
uses $\log L_X$ for the nuclear luminosity.
From the optical PCA performed on this sample of 42 objects, 
we find that 96.1\% of the variance can be explained with just the first two
eigenvectors ($\bvec{e_1}$ and $\bvec{e_2}$), and the QSOs mostly lie in a plane within this parameter
space.  This we consider to be a fundamental plane for QSOs.  
For the corresponding x-ray results, the first two eigenvectors explain 95.2\% of the variance in the sample, and we find here an x-ray QSO fundamental plane.
See Table~\ref{tab:pca-overall-optandx-42} for the individual eigenvalues.

\begin{figure}
\plottwo{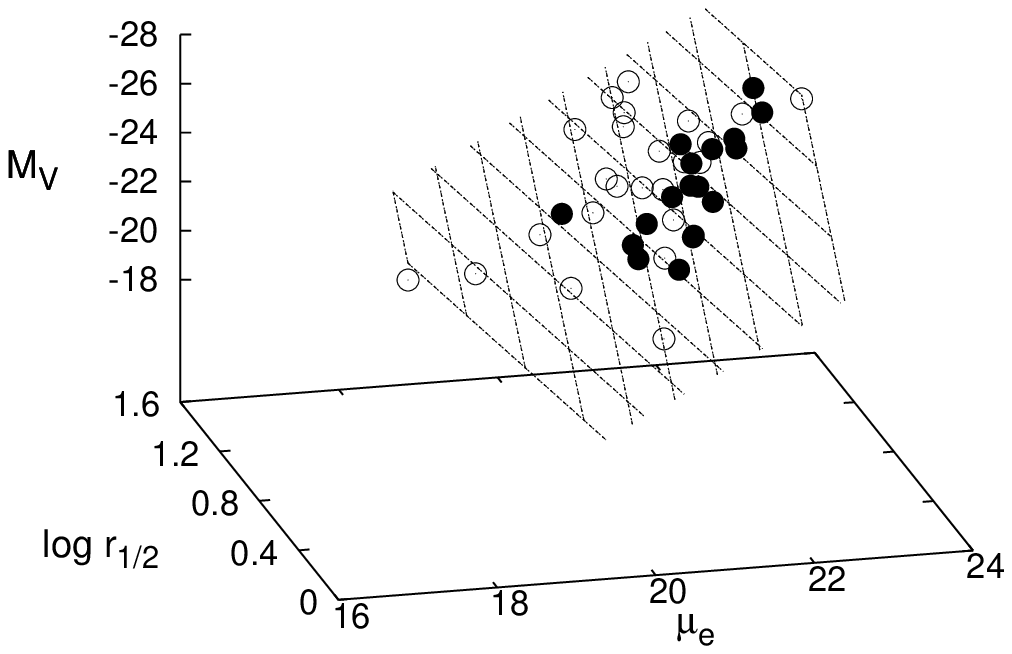}{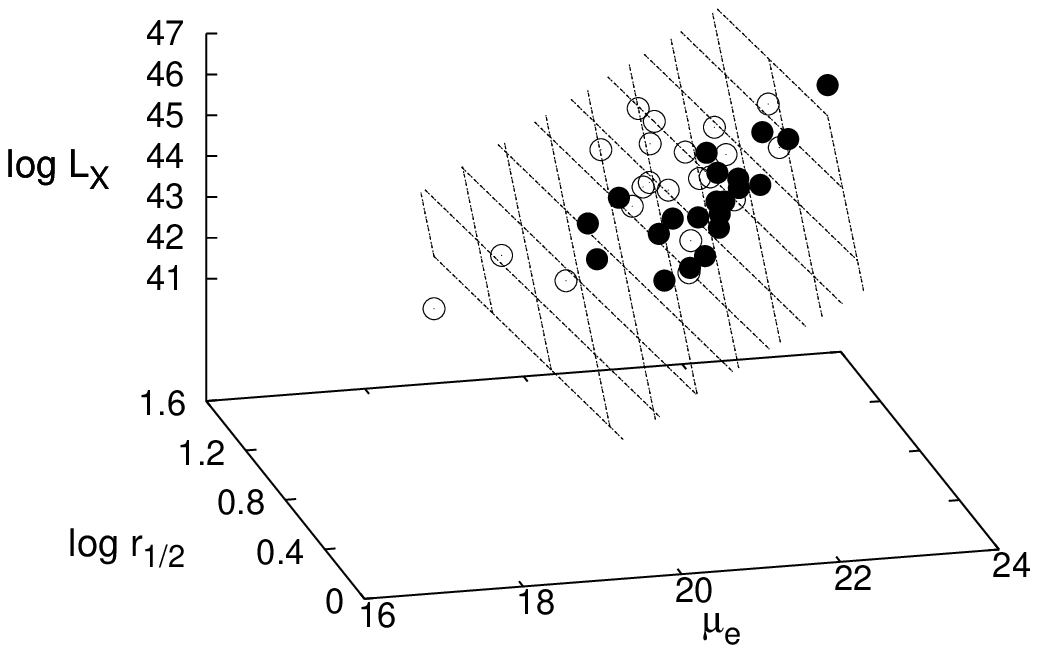}
\caption{
Views of the optical (left) and x-ray (right) QSO
fundamental planes, showing the individual QSOs (filled and open circles) and the plane (grid) fitted to the overall sample.  
The host properties are the horizontal axes, while nuclear luminosity is vertical.   In both cases, nuclear luminosity increases upward.  
The plots are viewed from above the plane; 
filled circles lie above the plane, and open circles lie below.  
}
\label{fig:fp-phys}
\end{figure}

We obtain the formula for the optical fundamental plane of the full sample,
\begin{equation}
	M_V\mathrm{(nuc)} = -77.5 + 3.14 \mu_e - 14.2 \log
	r_{1/2} \mbox{ .} \label{equ:oall-physical}
\end{equation}
The x-ray fundamental plane for the full sample is
\begin{equation}
	\log L_X = 79.3 - 2.03 \mu_e + 8.74 \log
	r_{1/2} \mbox{ .} \label{equ:xall-physical}
\end{equation}
Views of the optical and x-ray fundamental planes, with the QSO data points superimposed, 
are displayed in Figure~\ref{fig:fp-phys}.  Note that the host properties describe the horizontal and the nuclear luminosity the vertical in this plot.

The same analysis can also be made of the separate QSO subsamples.  We save the mathematics and analysis details for \S\ref{sec:appendix-b} but discuss the results in more general terms in \S\ref{sec:disc-fp} and \S\ref{sec:disc-fp-final}, below.  We find that, indeed, the subsamples are individually distributed into planes, with their first two eigenvectors accounting for about as much variance as in the full sample.  Some of the subsample planes, however, differ in orientation.

\subsubsection{Precision}

The precision of the fundamental plane can be judged from the
percentage of the variance explained by the third eigenvector, as listed in
Table~\ref{tab:pca-overall-optandx-42}.  
The third eigenvector describes the scatter perpendicular to the plane.

We can obtain another measure of the fundamental
plane's errors by comparing the measured nuclear luminosities with the values we
calculate from the QSO FP.  This can be done when solving for any of the
FP's three variables in terms of the other two.  The Pearson 
correlation coefficients and the root-mean-squared
differences between the actual values and their fundamental plane
solutions are listed in Table~\ref{tab:fp-rms}.  
The Pearson correlation coefficient is useful for accounting for the
size of the error, relative to the range of the variable.  

\begin{figure}
\plottwo{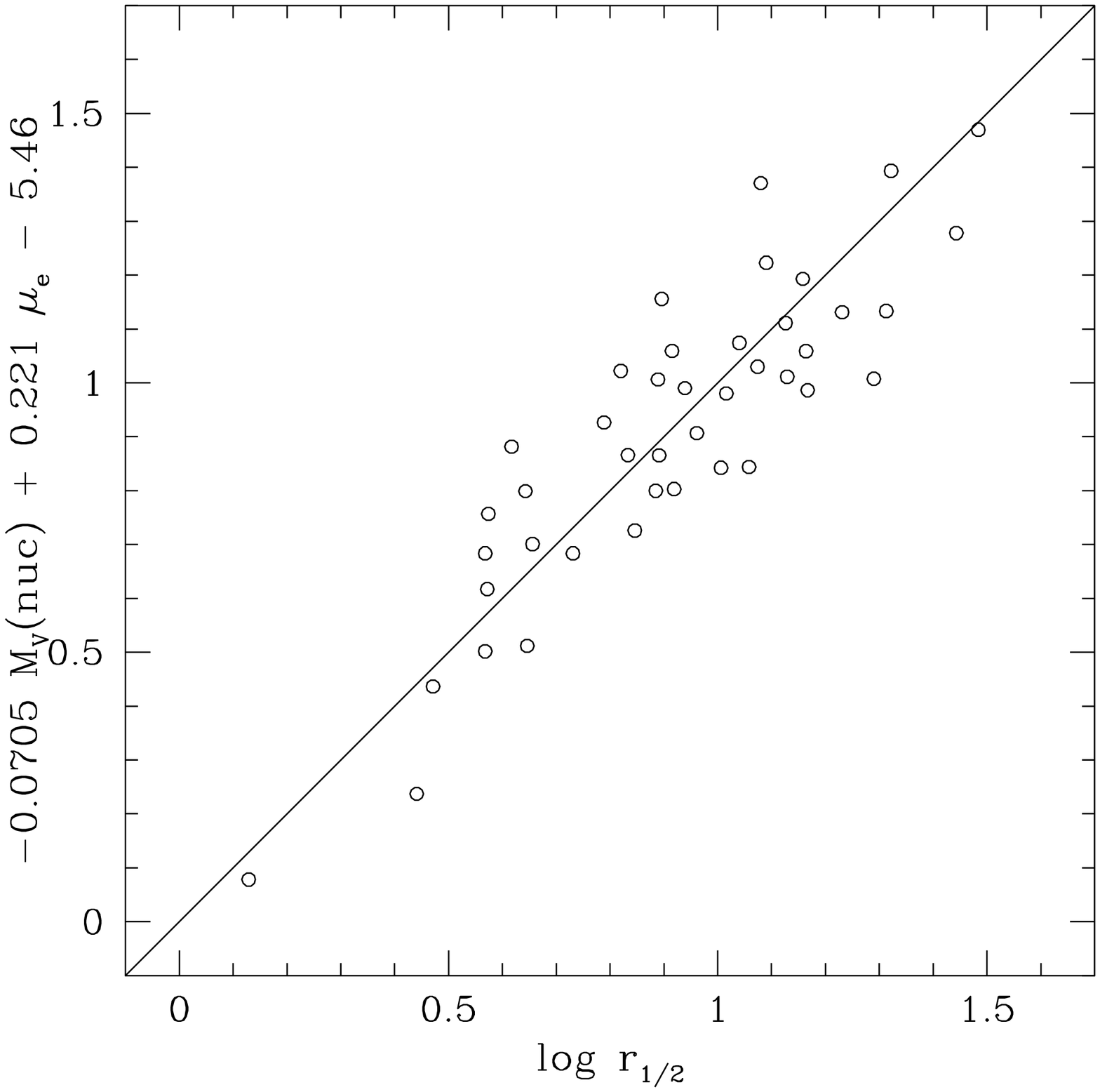}{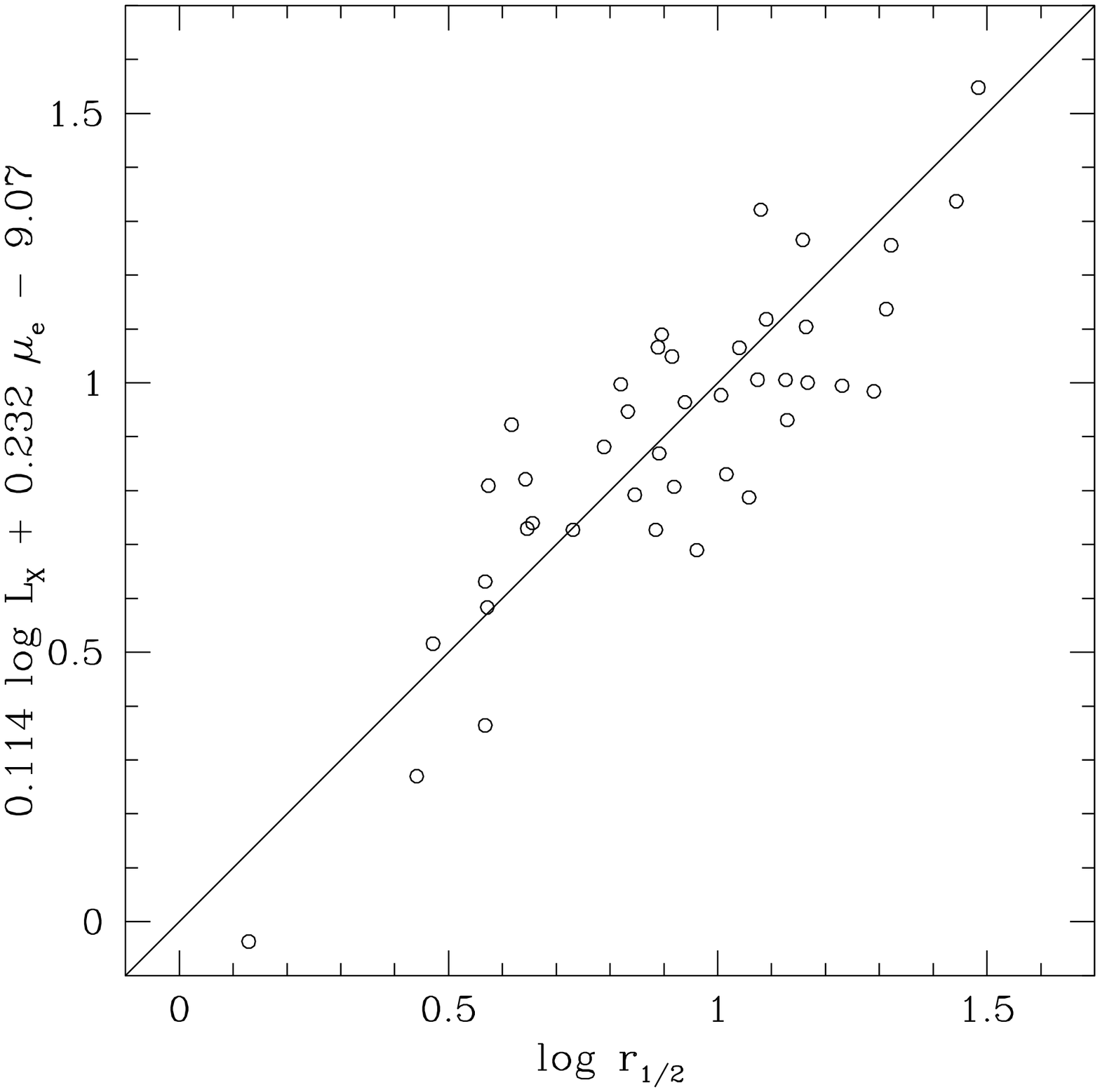}
\caption{Overall QSO fundamental plane (vertical
axis), plotted against the measured host galaxy size ($\log r_{1/2}$, horizontal
axis).  Points on the diagonal line show perfect correspondence.  
The left figure uses the QSO fundamental plane in its optical form, while the right
figure uses the x-ray form.  The QSO fundamental plane is most precise when solved for the host size.}
\label{fig:fp-rms}
\end{figure}

It is clear from these results that the QSO FP has both the smallest
root-mean-square (RMS) errors and the highest correlation with the data
when it is solved for the host galaxy size, $\log r_{1/2}$.  The full-sample 
fundamental plane in this form is
\begin{equation}
	\log r_{1/2} =  - 0.0704 M_V\mathrm{(nuc)} 
	+ 0.221 \mu_e -5.46
\label{equ:qso-ofp-final}
\end{equation}
for the optical, or 
\begin{equation}
	\log r_{1/2} =  0.114 \log L_X 
	+ 0.232 \mu_e -9.07
\label{equ:qso-xfp-final}
\end{equation}
for the x-ray.  The measured host sizes are plotted against the
fundamental plane in Figure~\ref{fig:fp-rms}.

\subsubsection{Effect of Statistical Errors}

The effect of the sample's statistical errors on the measurement of the QSO fundamental plane can be tested by creating an artificial FP, adding errors, and trying to recover the FP using the PCA method described above.  In this test, we first create a fake QSO FP equation.  We then randomly place 42 objects within the limits of $\log r_{1/2}$ and $\mu_e$ covered by the 42 real QSOs of the PCA sample.  Each object's nuclear luminosity ($M_V\mathrm{(nuc)}$ or $\log L_X$) is calculated from the fake FP equation.  We next add errors to each data point in the following way.  

The measurement uncertainties for each QSO are listed in Table~\ref{tab:pcasample}, and we use them to throw the errors.  For each object in the fake FP, we use the uncertainties from a different QSO of the actual sample.  The error for a given parameter is thrown randomly, according to a two-sided Gaussian of width equal to the QSO's uncertainty.  The parameter value for the fake object is then shifted by this random error.  In this way, the fake QSO FP is scrambled to simulate measurement errors.

We then run a Principal Components Analysis on the fake, scrambled FP.  After we derive the equation for it, we compare this with the equation we used to create the fake dataset in the first place.  Using this technique, we not only test the actual optical and x-ray QSO FPs, but we also vary their coefficients.  This lets us find out whether the statistical errors push the measured QSO FP into a particular direction.  Starting from the QSO FPs in the form of equations~(\ref{equ:oall-physical}) and (\ref{equ:xall-physical}), we individually vary the coefficients of $\log r_{1/2}$ and $\mu_e$, replacing them with values 0.5, 1, and 2 times their actual amounts.  It turns out that the statistical errors do not greatly affect the measured angle of the QSO FP, regardless of how we vary the equation's coefficients.  The coefficients ``recovered'' from the scrambled data are always within 2\% of their assigned values and in most cases are within 1\%.

\subsection{Discussion}\label{sec:disc-fp}

The existence of the QSO fundamental plane shows that there is a strong link between a 
QSO's host and nuclear properties, although we cannot make a claim as to cause and effect.
We find that the QSO subsamples have their own fundamental planes and that there is a 
great difference in their slopes (we specifically refer to the gradient).  
Mathematically, the FP gradients calculated in \S\ref{sec:appendix-b} 
describe the size of the dependence of 
the nuclear luminosity on the features of the host.  A large gradient, as in the x-ray form of the radio-quiet sample, means that a small change 
in the host size or effective surface magnitude is associated with a large change in the nuclear luminosity.  
Because the FP is a function of two parameters, looking only at the gradient obscures the details of 
this relationship somewhat.
But it is useful in that it demonstrates the underlying similarity between the 
FPs of the different subsamples, a similarity that is not obvious from looking at the individual 
equations.
Geometrically, the different planes act almost as a single plane that pivots about its principal axis, 
changing its angle relative to the host $(\log r_{1/2}, \mu_e)$ 
plane.  That is, the gradients of most of the subsamples point along the same 
(or opposite) azimuth, and the major difference between their FPs is the magnitude of their gradients.
More interestingly, not only does the gradient change magnitude, but it even reverses direction in some cases.  
In the optical FPs, only the QE sample slopes in the direction opposite the rest.  In the x-ray FPs, all of the radio-quiet 
samples (Q, QE, and QS) exhibit this behavior.  In fact, if we rank the planes by their gradients, 
radio-loudness seems to have the biggest influence in how the subsamples are grouped.

The statistical errors in the final PCA dataset do not appear to affect the 
measured orientation of the QSO FP significantly.  While this does not itself discount 
the possibility that systematic errors could have an effect, 
we note the difference between the FPs of the radio-loud ellipticals (LE) and radio-quiet 
ellipticals (QE).  These two FPs have slopes of opposite sign, 
but they deal with hosts of the same morphology.  Thus systematic errors from the 
image fitting procedure are unlikely to create the QSO FPs artificially.  While they 
might still influence the measured slope, they are not enough to cover up these 
large differences between the subsample FPs.  In addition to statistical and modeling errors, 
the nuclear luminosity is affected by QSO variability.  This is another source of FP thickness.

Finally, we have adopted a Friedmann cosmology with $H_0=50$ km
s$^{-1}$ Mpc$^{-1}$, $q_0=0.5$, and $\Lambda = 0$ to be in keeping
with Hamilton~(2001) and Hamilton et al.~(2002).  A
more currently-accepted cosmology might have $H_0 = 65$ km
s$^{-1}$ Mpc$^{-1}$, $\Omega_{\mathrm{matter}} =0.3$, and
$\Omega_{\Lambda} =0.7$.  The fundamental plane variables can be
converted into this cosmology for each individual QSO, but the
inclusion of a non-zero cosmological constant makes the
conversions depend on redshift.

The differences are smallest for QSOs at the high-redshift end of the
sample.  Converting to the updated cosmology lowers the
values of $\log r_{1/2}$ by 0.101, for QSOs at the low-redshift limit of $z=0.06$, 
and it lowers them by 0.039, for QSOs at the high-redshift limit of $z=0.46$.  
The conversion makes the magnitudes,
$M_V{\mathrm{(nuc)}}$ and $\mu_e$, fainter by 0.505 to
0.195 mag, and it lowers $\log L_X$ by 0.202 to 0.007 (again, the 
low-redshift changes are listed first).  
These changes are small compared to the ranges of the variables.
More importantly, they do not produce any significant effect in the 
QSO fundamental plane quality.  In the optical case, the percentage of the 
sample variance explained by the PCA would be lowered from 96.1\% to 95.6\%, 
and in the x-ray case, from 95.2\% to 94.7\%.

%%%%%%%%%%%%%%%%%%%%%%%%%%%%%%%%%%%%%%%%%%%%%%%%%%%%%%%%%%%%%%%%%%%%%%

\section{DISCUSSION}\label{sec:discussion}

\subsection{Host and Nuclear Luminosity}\label{sec:finaldisc-hostnuclum}

We have shown that host luminosity shows a shallow increase with
nuclear luminosity.  Some relationship of this type is expected because in more massive 
(and therefore more luminous) hosts, there is
more gas available to fuel the supermassive black hole thought
necessary to power a QSO.  

Bahcall et al.~(1997) do not
find convincing evidence in their results of a significant correlation
between host and nuclear luminosities, and they do not provide a
fitted relation, noting that much of the apparent correlation in their
Figure~8 comes from 3C~273, the brightest object in their sample.

O'Dowd et al.~(2002) find a shallow but statistically
significant trend of host luminosity increasing with nuclear
luminosity.  For the combined sample, host luminosity increases by 1
mag or less (depending on beaming corrections for the BL Lacs) for
each 7 mag increase in nuclear luminosity.  Their
study uses a mixed sample of objects (BL Lacs and QSOs).

For their radio-loud QSOs, Dunlop et al.~(2003) find a shallow
relationship, $M_R\mathrm{(host)} = +0.114 (\pm 0.15)
M_R\mathrm{(nuc)} -20.98 (\pm 3.72)$.  For radio-quiet QSOs, they find
a nearly-flat relationship, $M_R\mathrm{(host)} = -0.018 (\pm 0.18)
M_R\mathrm{(nuc)} -23.73 (\pm 4.29)$, when they reject the two
lowest-luminosity objects as not being true QSOs.  Overall, they
report that both results are consistent with there being no
correlation, using the Spearman rank correlation test.

\subsection{Bulge Luminosity and Black Hole Mass}\label{sec:finaldisc-bulge-bh}

Without interpreting the weak bulge and black hole 
result too far, we note for comparison that our fitted relation, 
equation~(\ref{equ:mbulge-mbh-orig}), is fairly consistent with 
that derived by McLure \& Dunlop~(2001).  It corresponds to 
their equation~(3), after transforming their $M_{R_C}$ to $M_V$ with the 
elliptical galaxy colors from Fukugita et al.~(1995).  
However, their newer results with an expanded sample (McLure \& Dunlop~2002) 
find a steeper slope than our equation~(\ref{equ:mbulge-mbh-orig}).   
We should also note that a reanalysis of their data by Graham~(2007), 
editing the sample and using a different 
linear regression method, steepens it to almost twice our slope.  
Graham finds close agreement between this and his similar 
reanalysis of Marconi \& Hunt~(2003).

Neither McLure \& Dunlop nor Marconi \& Hunt restrict themselves only 
to QSOs, and so they have samples extending to much 
lower black hole masses than ours, which improves their correlations. 
The problem with looking at a QSO population alone, as we do, is that the short 
range of masses is not much greater than the measurement errors.  
More data are needed for QSOs with less massive black holes, if true
QSOs even exist with much smaller masses.  
And the real test will come when the uncertainties in black hole 
mass measurements have been reduced.

\subsection{Nuclear Luminosities and Eddington Fractions}\label{sec:finaldisc-nuc-edd}

We see that nuclear luminosity generally increases with black hole
mass, although the trend is not strong.  Some kind of relationship might be
expected, since the gravity of the black hole is the ultimate energy
source for the QSO.  For instance, as black hole mass increases, more material can
be trapped by its gravity and be pulled into the accretion disk; the
faster material accretes onto the disk, the more the disk heats up and
light is produced.  But any effect produced by this is rather weak.

We can see some trends, though, when we look at the Eddington fractions 
of the QSOs.  The QSOs in spirals tend to be radiating at higher Eddington 
fractions than those in ellipticals.  While spirals generally have lower nuclear 
luminosities, they also tend to have smaller black holes.  
This could be a result of the amount of gas available to fuel 
the QSO.  If there is a fixed rate of gas flowing into the vicinity of
the black hole, owing to the shape of orbits in the galactic bulge, then
it may be that a smaller black hole, with a lower Eddington limit, is
able to capture enough material to keep it radiating near its limit.
But as the black hole grows, although it is able to capture more of
the material orbiting near it, there is not enough available to keep
it near the increased Eddington limit.  
It could also simply be that if, rather than limiting ourselves to 
traditionally-defined QSOs, we were to include lower-luminosity Seyfert galaxies in our sample, we might 
find them populating the low Eddington fraction region of the spiral host plot.

A criticism of one of the trends described above 
comes from Woo \& Urry~(2002), who suggest that what appears to
be a correlation between black hole mass and nuclear luminosity  
might be explained by sample selection effects.  
They argue that there is not sufficient evidence
for a correlation, 
because the lack of objects in the lower right side of
Figure~\ref{fig:mnuc-mbh} is due to the exclusion of lower-luminosity 
AGN from the sample.  They state that radio galaxies, for
instance, can have low nuclear luminosity but high black hole mass.
The exclusion from our sample of radio galaxies might
not have exactly the same effect on our distributions as it does for
Woo \& Urry~(2002), since they include radio galaxies but are not able
to obtain good bolometric measurements for the least
luminous ones.
The upper left side of the plot is real, however, because of the
Eddington limit.  

Finally, we should also say that there is an added uncertainty from QSO variability 
and the possible beaming of the nuclear emission.  
For instance, PG~1302-102 is variable (Eggers, Shaffer, \& Weistrop~2000 
show changes of about 0.2 mag), and its correlated optical and radio structure 
could indicate beaming (Hutchings et al.~1994b).  
There should be little beaming among the radio-quiets.  
But among the radio-louds, it could exaggerate some nuclear luminosities  
and Eddington fractions.

\subsection{Fundamental Plane}\label{sec:disc-fp-final}

The fundamental plane for QSOs shows a relationship between the
nuclear and host features that goes beyond the simple correlation of nuclear and
host luminosities.  This behavior may be connected to other, known
relations between the objects, such as the fundamental plane of
normal, elliptical galaxies.  The fundamental plane for elliptical
galaxies involves the galaxy's size, effective surface magnitude, and
stellar velocity dispersion.  It varies quantitatively for galaxies in different environments and for
observations in different wavebands, but one measurement in the 
$V$-band (Scodeggio et al.~1998) is
\begin{equation}
    \log r_{1/2} = 1.35 \log \sigma_c + 0.35 \mu_e + \mathit{Constant ,}
\label{equ:ellip-fp}
\end{equation}
where $\sigma_c$ is the galaxy's central velocity dispersion.

Although the ratio of the coefficients of the half-light radius to the
effective surface magnitude is different from the QSO fundamental plane, 
there is a formal similarity between 
the QSO and normal FPs, which might point to a link between 
the host galaxy's central velocity
dispersion and the nuclear luminosity of the QSO.  
It is therefore tempting to try to derive the
QSO fundamental plane directly from the elliptical galaxy fundamental
plane, but we find two problems with this approach, both arising from the relation of black hole mass to nuclear luminosity.  

By applying the correlation between the central velocity dispersion 
in elliptical galaxies and the mass of their central black hole, 
we can put the elliptical galaxy fundamental plane,
equation~(\ref{equ:ellip-fp}), in terms of black hole mass.  
We use the
formula of Merritt \& Ferrarese~(2001),
\begin{equation}
	\mathcal{M}_{BH}=1.3 \times 10^8 
	\left( \frac{\sigma_c}{200 \mbox{ km s}^{-1}} \right)^{4.72} 
	\mathcal{M}_{\sun}\mathrm{ .}
\label{equ:sigma-mbh}
\end{equation}
Further substituting in equation~(\ref{equ:mnuc-mbh}) to put this in terms of
nuclear luminosity, we obtain
\begin{equation}
	\log r_{1/2} = - 0.14 M_V\mathrm{(nuc)} + 0.35 \mu_e + \mathit{Constant .}
\label{equ:qso-ofp-backderived}
\end{equation}
This is our attempt to derive the QSO optical fundamental plane from the normal galaxy FP, and 
we see that it differs from the actual QSO FP, equation~(\ref{equ:qso-ofp-final}).

The coefficient we derive for
$M_V\mathrm{(nuc)}$ in the above equation depends upon the exponent of the velocity
dispersion in equation~(\ref{equ:sigma-mbh}) and on the slope of
equation~(\ref{equ:mnuc-mbh}).  Different studies have found exponents for the velocity
dispersion (a hotly debated topic) that range from about 3.75---5.1 
(Ferrarese \& Merritt~2000; Gebhardt et al.~2000; Tremaine et al.~2002; Ferrarese \& Ford~2005; Bernardi et al.~2007), 
and the slope of equation~(\ref{equ:mnuc-mbh}) has a large uncertainty.
But we still cannot make the 
coefficient of $M_V\mathrm{(nuc)}$ in equation~(\ref{equ:qso-ofp-backderived}) 
match that of the QSO optical fundamental plane.  
Deriving the QSO fundamental plane directly from the 
normal galaxy FP would require a steep dependence of $M_V\mathrm{(nuc)}$ upon 
$\log \left( \mathcal{M}_{\mathrm{BH}} / \mathcal{M}_{\odot} \right)$, which we do not find.

We can attempt to derive the x-ray QSO fundamental plane the same way, using the 
correlation between the x-ray luminosities and black hole masses in our sample.  
A linear fit using the BES method (in keeping with the derivation 
for the optical plane) gives us 
$\log L_X = 2.77 \log \left( \mathcal{M}_{\mathrm{BH}} / \mathcal{M}_{\odot} \right) + 19.8$.  
This would lead to 
\begin{equation}
	\log r_{1/2} = 0.10 \log L_X + 0.35 \mu_e + \mathit{Constant}
\end{equation}
as our attempted derivation, which is close to the actual x-ray QSO FP.
The use of x-ray luminosities normalized to 0.5 keV glosses over the complexities of 
QSO spectra, providing only a first glimpse of the x-ray QSO fundamental plane.  Follow-up analysis 
should account for the different spectral features of the individual QSOs 
and should use luminosities of a higher energy band.

There is both an interesting property and a problem in the QSO FP orientations, which 
we detail in \S\ref{sec:appendix-b}.  
As we change from one QSO class to another, the fundamental plane 
essentially pivots about an axis, so the differences between the subsamples 
are mostly reduced to a single dimension, 
the slope (or gradient) relative to the $\mu_e$--$\log r_{1/2}$ plane.  
Some subsample FPs even slope in the opposite direction from the overall FP.  
This creates another problem for deriving the QSO FPs (whether optical or x-ray) from the 
normal galaxy fundamental plane.  These opposite slopes 
are not reflected in our fits of nuclear luminosity 
against black hole mass, and the poor correlation of these two 
quantities lies in contrast with the relatively thin QSO fundamental plane.  
This points to our need for a better understanding of the proper 
relationship between nuclear luminosity and black hole mass.

Because the QSO FP mathematically describes a link between the host and the 
nucleus, its slope might depend on the physical nature of the link.  
It would be interesting to find if the different QSO FP orientations described in \S\ref{sec:appendix-b} come 
about from different fueling mechanisms that might be found in the various subsamples.
We see, for instance, that radio-quiet and radio-loud QSOs are characterized by very 
different slopes in their x-ray FPs, but the understanding of what makes these QSO types differ 
is still too limited to speculate further here.  In our ongoing research, we are expanding the 
fundamental plane study to other types of AGN, such as Seyferts.  
We can then compare their FP orientations with those of the different QSO
subsamples, which may teach us more about the physics underlying the QSO fundamental plane.

\section{CONCLUSIONS}\label{sec:conclusions}

1. QSO hosts cover a much smaller range of luminosity than do their
   nuclei.  We find a relatively weak trend between host and nuclear
   luminosities.

2. Host bulge luminosities do not show a significant correlation with
   H$\beta$-derived black hole masses, but a Magorrian-type relationship 
   cannot be ruled out.

3. The correlations seen among nuclear luminosity, black hole mass,
   and Eddington fraction are subject to selection effects from the
   exclusion of lower-luminosity AGN.

4. We show that there is a strong, 3-parameter 
   relationship between host and nuclear properties.  The
   nuclear luminosity, size of the host, and effective surface
   magnitude of the host together form a plane in which most of the
   QSOs lie.  This fundamental plane for QSOs explains 96.1\% of the
   variance in the sample, in its optical form, and 95.2\% in its
   x-ray form.  While this plane shows some similarities to the
   fundamental plane for elliptical galaxies, we do not find a clear,
   analytical derivation from one to the other.

\acknowledgements

This research has been supported through a National Research Council
Associateship at NASA's Goddard Space Flight Center, as well as
through a Graduate Student Fellowship and a grant from the Director's
Discretionary Research Fund, both from the Space Telescope Science
Institute.  We made use of the following databases: the HST data
archive; NASA/IPAC Extragalactic Database (NED), operated by the Jet
Propulsion Laboratory, California Institute of Technology, under
contract with the National Aeronautics and Space Administration; and NASA's
Astrophysics Data System Abstract Service.  We would like to thank
George Djorgovski for his comments on the interpretation of the
fundamental plane.  We are also grateful for the long support and 
encouragement of the late Elizabeth K. Holmes, a National Research Council 
Associate at JPL.

%%%%%%%%%%%%%
%% APPENDIX A %%
%%%%%%%%%%%%%

\appendix

\section{COMPARISON WITH OTHER QSO IMAGING RESULTS}\label{sec:appendix-a}

An idea of the systematic differences between our analysis method and others is gotten by comparing our QSO imaging results with those of other researchers using the very same data sets (Table~\ref{table:complist2}).  The most useful comparisons are of the nuclear and host apparent magnitudes, as well as the host effective radius.  We include only those results that can be directly compared with ours--those reporting the apparent WFPC2 filter magnitudes in the Vega magnitude system.  

As discussed in \S\ref{sec:imageanalysis} and more extensively by Hamilton et al.~(2002), we first match the focus and the PSF sub-pixel center and then apply a simultaneous nucleus and host fit.  Our host magnitudes are measured from the PSF-subtracted image, with the host model used to fill in masked pixels and to extrapolate beyond the aperture to infinite radius.  
There are several objects, flagged in Table~\ref{table:complist2}, whose host sizes are difficult to compare, 
but which we have included nonetheless.  
These are generally spiral hosts with both a bulge (or bar) and a disk, 
which the literature analysis treats as having only a single component.  
We either model both the bulge and the disk, or we mask one component and fit only the other.  
In either case, our magnitudes reflect 
the total host magnitude, but the radius (taken directly from the model) represents only one component.  
See the table notes for the details.

Next are particular comparisons with the groups with which we share several objects.
Closely matching our results are those of McLure et al.~(1999), which are often within 0.1 mag of ours.  They use a similar technique, a simultaneous nucleus and host fit to the 2-D image, that differs from ours in the specifics of the PSF and host models.  The close correspondence of our results argues for a robustness in the approach.
The size scales of our host galaxies are, on average, about 1.2 ($\pm 0.3$) times larger than those fitted by McLure et al.~(1999), 
ignoring those objects flagged in Table~\ref{table:complist2} where we have fitted models to different components of the host.  
Note that although the apparent magnitudes are listed as Cousins $R$, they are, in fact, in the $F675W$ filter.  
McLure says that the $R_C - F675W$ color conversions for these objects turned out to be too small to make a difference.

Larger differences, on the other hand, can be seen between our results and those of Bahcall et al.~(1997) and Boyce et al.~(1998).  Their analysis methods are quite different from ours.  
Bahcall et al.~(1997) progressively subtract a stellar nucleus until the residual profile turns over in the center, then fit a host model to the remaining light.  We compare with their 2-D model results.  
In principle, this method should somewhat overestimate the host magnitude.  
Boyce et al.~(1998) fit simultaneous nucleus and host models using the cross-correlation method described by Boyce, Phillipps, \& Davies~(1993).  Host magnitudes are measured directly from the light remaining after PSF subtraction.   All of their images are marred by saturation in the innermost pixels, which they remove entirely from the analysis and set to null for the host magnitude calculation.  Thus their host magnitudes are upper limits, as they state.  

Lastly, Hooper et al.~(1997) use the same cross-correlation method as Boyce et al.~(1998), although they apparently take the host magnitudes from the models, rather than from the PSF-subtracted images.  They report their results in the Johnson $R$ filter, rather than the WFPC2 F675W in which they were taken (stated to be a $< 0.15$ mag conversion), and they apply other magnitude corrections ($\leq 0.1$ mag).  The conversions are color-dependent, and the corrections are not explicitly given, so we have left these in Johnson $R$.  Sizeable differences with our results can be seen, but the magnitude differences should be treated as rough.

%%%%%%%%%%%%%
%% APPENDIX B %%
%%%%%%%%%%%%%

\section{SUBSAMPLE FUNDAMENTAL PLANES}\label{sec:appendix-b}

\subsection{Subsample Forms}\label{sec:fp-forms}

In addition to the full sample, 
we also look for fundamental planes in the individual QSO subsamples; their PCA 
eigenvalues are presented in Table~\ref{tab:pca-comprehensive-42}.  As before, the subsamples are labeled
``L'' for radio-loud QSOs, ``Q'' for radio-quiet, ``E'' for those in elliptical
hosts, and ``S'' for those in spirals.  Because the restricted sample
includes only a single radio-loud QSO in a spiral host, there is no
separate FP for the ``LS'' subsample.

We find that, indeed, the subsamples are individually distributed into planes, and the third eigenvalue (representing scatter perpendicular to the plane) in each case is about as small as the full sample's.  Table~\ref{tab:fp-physicalforms} lists the coefficients for the subsamples' fundamental plane equations.  The equations are presented in the optical case as $M_V\mathrm{(nuc)}=R_{\mathrm{coeff}} \log r_{1/2}+M_{\mathrm{coeff}} \mu_e+C$ and in the x-ray case as $\log L_X=R_{\mathrm{coeff}} \log r_{1/2}+ M_{\mathrm{coeff}} \mu_e+C$, where $R_{\mathrm{coeff}}$ and $M_{\mathrm{coeff}}$ are coefficients, and $C$ is a constant.  The orientations of some of these planes are clearly different from the others.  

\subsection{Gradients}

The situation is more easily described geometrically, because it turns out that their biggest differences can be characterized by a single parameter, the slope (magnitude of the gradient).  
To visualize the relationship among the planes, 
let us imagine a 3-D plot similar to Figure~\ref{fig:fp-phys}, 
but rescaled, replacing $M_V\mathrm{(nuc)}$ with $M_V\mathrm{(nuc)}/2.5$ and $\mu_e$ with 
$\mu_e/2.5$, so that all quantities are pure logarithms.  
This will allow a more straightforward comparison of lengths and angles.  
For the vertical axis ($\bvec{k}$), we will use the nuclear 
luminosity ($\log L_X$ or $M_V\mathrm{(nuc)} / 2.5$), the $\bvec{i}$ axis will be 
$\log r_{1/2}$, and the $\bvec{j}$ axis will be $\mu_e / 2.5$.  
Again, the host properties describe the horizontal and the nuclear luminosity the vertical.  

In the following, let us represent the rescaled QSO fundamental plane by the function 
$F = F (\log r_{1/2}, \mu_e/2.5)$.  
The gradient of the fundamental plane  
is $\nabla F = \left(\partial F/\partial \log r_{1/2} \right) \bvec{\hat{\imath}} 
+ \left[\partial F/\partial (\mu_e/2.5) \right] \bvec{\hat{\jmath}}$, 
where $\bvec{\hat{\imath}}$ and $\bvec{\hat{\jmath}}$ are unit vectors.
Note that since $F$ is a plane, its gradient is the same at any point we choose to evaluate it.  
We can look at the magnitude and direction of the gradient, where the magnitude is given by 
\begin{equation}
	\left| \left| \nabla F \right| \right| = 
	\left\{ \left( \frac{\partial F}{\partial \log r_{1/2}} \right) ^2
	+ \left[ \frac{\partial F}{\partial \left( \mu_e/2.5 \right)} \right]^2 \right\}^{1/2} \mathrm{ ,}
\end{equation}
and the direction (azimuth, the angle counterclockwise from the $+\bvec{i}$ axis) is given by 
\begin{equation}
	\alpha = \arctan \left[ \frac{\partial F / \partial \left( \mu_e/2.5 \right) } 
	{ \partial F / \partial \log r_{1/2} } \right] \mathrm{ .}
\end{equation}

These gradients are listed in Table~\ref{tab:fp-gradients}.  The full sample is listed first, while the subsamples are grouped according to general azimuth and then ranked in order of gradient magnitude.  
The subsample gradients tend to point either along the same azimuth as the overall FP or in the opposite direction ($180^{\circ}$ away from it).  This near-uniformity in azimuth lies in contrast with the wide range of magnitudes (from 6.45 to 65.9), and so the gradient magnitudes seem to be the distinguising feature between the subsample planes.  

It should not be surprising that the azimuths are nearly uniform, as the first principal axis of the PCA follows roughly the Kormendy relation (Kormendy~1977) between $\mu_e$ and $\log r_{1/2}$.  What is interesting, then, is the large degree of freedom for the planes about this axis.

\subsection{Discussion}
The x-ray FPs have an interesting symmetry in their gradients.  
We see the the L and Q classes are paired with each 
other in having the largest gradients (except for the S class, discussed below) 
but pointing in opposite directions.  
This opposition in gradient directions is followed by the mixed classes, with the 
LE plane on one side and the QE and QS planes on the other.  
Note that the LS class is too small to derive a separate fundamental plane for it.
The S class is an exception to this symmetry.  For the PCA sample, the only difference 
between the S and QS populations is 
a single object, 3C~351, which is radio-loud and apparently spiral.  
With the small populations of these two classes, this one QSO makes a real difference.
In general, we find that the various planes for radio-loud objects point in one direction, 
while those for radio-quiets point in the opposite direction. 

The optical FPs all have approximately the same azimuth, with the exception 
of the QE class, which points in the opposite direction.  
But even among the optical planes, there are distinct groupings 
(mostly by radio-loudness) that almost mirror those of the x-ray FPs.
Aside from the QE class, the spirals and radio-quiet objects have the largest gradients, 
while the ellipticals and radio-loud objects have the smallest.
It seems that radio loudness makes the greatest difference in the gradient and azimuth, 
both in the x-ray and the optical FPs, while the host morphology appears to be less important.

\clearpage

%%%%%%%%%%%%%
%% FIGURES %%
%%%%%%%%%%%%%

\clearpage

\clearpage

\clearpage

\clearpage

\clearpage

\clearpage

%%%%%%%%%%%%
%% TABLES %%
%%%%%%%%%%%%

\begin{deluxetable}{llccccccc}
 \tabletypesize {\tiny}
\setlength{\tabcolsep}{0.05in}
 \tablewidth{0pt}

\tablecaption{Observations and Data}

\tablehead{
\colhead{$\begin{array}[t]{c} \mbox{Name} \\ (1) \end{array}$}&
\colhead{$\begin{array}[t]{c} z  \\ (2)  \end{array}$}&
\colhead{$\begin{array}[t]{c} m_{\mathrm{nuc}}   \\ (3)  \end{array}$}&
\colhead{$\begin{array}[t]{c} m_{\mathrm{host}}  \\ (4)  \end{array}$}&
\colhead{$\begin{array}[t]{c} M_V\mathrm{(nuc)}  \\ (5)  \end{array}$}&
\colhead{$\begin{array}[t]{c} M_V\mathrm{(host)} \\ (6)  \end{array}$}&
\colhead{$\begin{array}[t]{c} \mbox{Morphology} \\ (7) \end{array}$}&
\colhead{$\begin{array}[t]{c} \mbox{Radio} \\ (8) \end{array}$}&
\colhead{$\begin{array}[t]{c} \mbox{Selection} \\ (9) \end{array}$}
}

\startdata

\objectname[LBQS 0020+0018]{LBQS 0020+0018}   &0.423   &19.30  &19.34  &-22.45   &-22.77  &E    &Q  &A  \\
\objectname[LBQS 0021-0301]{LBQS 0021-0301}   &0.422   &19.03  &19.10  &-22.74   &-23.03  &E    &Q  &A  \\
\objectname[PG 0043+039]{PG 0043+039}      &0.385   &16.04  &19.03  &-25.49   &-22.67  &E    &Q  &B  \\
\objectname[PG 0052+251]{PG 0052+251}      &0.155   &16.04  &16.83  &-23.80   &-23.08  &S    &Q  &   \\
\objectname[PHL 909]{PHL 909}            &0.171   &15.97  &16.89  &-24.07   &-23.29  &E    &Q  &   \\
\objectname[UM 301]{UM 301}             &0.393   &17.66  &19.44  &-23.92   &-22.44  &E    &Q  &A  \\
\objectname[3C 47]{3C 47}              &0.425   &17.82  &18.75  &-24.07   &-23.37  &E    &L  &   \\
\objectname[3C 48]{3C 48}              &0.367   &15.74  &16.18  &-25.72   &-24.83  &EI   &L  &   \\
\objectname[PHL 1093]{PHL 1093}           &0.26    &17.21  &17.17  &-23.47   &-23.54  &E    &L  &B  \\
\objectname[MRK 1014]{MRK 1014}           &0.163   &16.17  &14.75  &-23.30   &-24.34  &S\tablenotemark{a}    &Q  &   \\
\objectname[PKS 0202-76]{PKS 0202-76}      &0.389   &16.67  &18.72  &-24.98   &-23.12  &E    &L  &   \\
\objectname[NAB 0205+02]{NAB 0205+02}      &0.155   &15.40  &18.08  &-24.37   &-21.77  &S    &Q  &   \\
\objectname[HB89 0244+194]{Q 0244+194}       &0.176   &16.80  &17.54  &-23.29   &-22.47  &E    &Q  &   \\
\objectname[US 3498]{US 3498}            &0.115   &19.30  &15.87  &-19.79   &-23.10  &S    &Q  &B  \\
\objectname[PKS 0312-77]{PKS 0312-77}      &0.223   &16.13  &16.67  &-24.40   &-23.78  &E    &L  &   \\
\objectname[HB89 0316-346]{Q 0316-346}       &0.260   &16.21  &18.08  &-24.68   &-23.03  &IS   &Q  &   \\
\objectname[3C 93]{3C 93}              &0.357   &18.58  &18.49  &-23.51   &-23.73  &E    &L  &   \\
\objectname[IRAS 04505-2958]{IR 0450-2958}     &0.286   &15.40  &17.17  &-25.41   &-23.65  &SI   &Q  &   \\
\objectname[PKS 0736+01]{PKS 0736+01}      &0.191   &16.30  &16.70  &-24.03   &-23.58  &E    &L  &   \\
\objectname[MS 07546+3928]{MS 07546+3928}    &0.096   &14.26  &14.37  &-24.22   &-23.47  &E    &Q  &   \\
\objectname[IRAS 07598+6508 NED02]{IR 0759+6508}     &0.149   &15.94  &15.65  &-23.57   &-23.73  &SI   &Q  &   \\
\objectname[MS 0801.9+2129]{MS 0801.9+2129}   &0.118   &16.00  &15.66  &-22.91   &-22.80  &S    &Q  &   \\
\objectname[3C 206]{3C 206}             &0.198   &16.07  &16.90  &-24.03   &-23.09  &E    &L  &   \\
\objectname[3C 206]{3C 215}             &0.412   &17.71  &18.23  &-24.08   &-23.08  &E    &L  &   \\
\objectname[PG 0923+201]{PG 0923+201}      &0.19    &15.53  &17.46  &-24.73   &-23.00  &E    &Q  &   \\
\objectname[MS 0944.1+1333]{MS 0944.1+1333}   &0.131   &14.89  &15.93  &-24.16   &-22.52  &E    &Q  &   \\
\objectname[PG 0953+414]{PG 0953+414}      &0.234   &15.17  &17.21  &-25.23   &-23.22  &S    &Q  &B  \\
\objectname[PG 1001+291]{PG 1001+291}      &0.330   &15.59  &17.90  &-25.58   &-23.33  &S    &Q  &B  \\
\objectname[PKS 1004+13]{PKS 1004+13}      &0.24    &15.15  &17.00  &-25.62   &-24.06  &E    &L  &   \\
\objectname[PG 1012+008]{PG 1012+008}      &0.185   &16.22  &16.76  &-23.75   &-23.19  &SI   &Q  &B  \\
\objectname[HE 1029-1401]{HE 1029-1401}     &0.086   &13.84  &15.86  &-24.79   &-22.74  &E    &Q  &B  \\
\objectname[MS 1059.0+7302]{MS 1059.0+7302}   &0.089   &16.60  &15.41  &-21.65   &-22.34  &S    &Q  &   \\
\objectname[PG 1116+215]{PG 1116+215}      &0.177   &14.85  &16.74  &-25.19   &-23.42  &S    &Q  &B  \\
\objectname[PG 1202+281]{PG 1202+281}      &0.165   &16.85  &17.39  &-23.03   &-22.62  &E    &Q  &B  \\
\objectname[LBQS 1209+1259]{LBQS 1209+1259}   &0.418   &19.35  &19.38  &-22.39   &-22.71  &E    &Q  &A  \\
\objectname[PG 1216+069]{PG 1216+069}      &0.331   &15.42  &18.70  &-25.76   &-22.55  &E    &Q  &   \\
\objectname[LBQS 1218+1734]{LBQS 1218+1734}   &0.444   &18.33  &19.01  &-23.54   &-23.26  &E    &L  &A  \\
\objectname[MS 1219.6+7535]{MS 1219.6+7535}   &0.071   &15.06  &14.56  &-22.72   &-22.52  &ED   &Q  &   \\
\objectname[LBQS 1222+1010]{LBQS 1222+1010}   &0.398   &18.38  &18.62  &-23.23   &-23.24  &S    &Q  &   \\
\objectname[LBQS 1222+1010]{LBQS 1222+1235}   &0.412   &17.68  &18.25  &-24.04   &-23.82  &E    &L  &A  \\
\objectname[3C 273]{3C 273}             &0.158   &12.60  &15.65  &-27.19   &-24.24  &E    &L  &   \\
\objectname[PG 1229+204]{PG 1229+204}      &0.064   &15.37  &15.04  &-22.27   &-22.33  &S    &Q  &C  \\
\objectname[LBQS 1240+1754]{LBQS 1240+1754}   &0.458   &17.98  &19.31  &-23.93   &-23.02  &E    &Q  &A  \\
\objectname[LBQS 1243+1701]{LBQS 1243+1701}   &0.459   &18.45  &18.44  &-23.49   &-23.91  &E    &Q  &A  \\
\objectname[3C 277.1]{3C 277.1}           &0.321   &17.97  &18.35  &-23.09   &-22.76  &S    &L  &   \\
\objectname[PG 1302-102]{PG 1302-102}      &0.278   &15.19  &17.35  &-25.93   &-24.14  &E    &L  &   \\
\objectname[PG 1307+085]{PG 1307+085}      &0.155   &15.46  &17.47  &-24.33   &-22.42  &E    &Q  &B  \\
\objectname[PG 1309+355]{PG 1309+355}      &0.184   &15.56  &16.61  &-24.53   &-23.62  &S    &L  &   \\
\objectname[PG 1358+04]{PG 1358+04}       &0.427   &15.96  &18.02  &-25.84   &-24.02  &E    &Q  &B  \\
\objectname[HB89 1402+436]{Q 1402+436}       &0.323   &15.15  &17.42  &-25.93   &-23.73  &EI   &Q  &   \\
\objectname[PG 1402+261]{PG 1402+261}      &0.164   &15.73  &17.33  &-24.12   &-22.62  &S    &Q  &   \\
\objectname[MS 1416.3-1257]{MS 1416.3-1257}   &0.129   &15.83  &16.90  &-23.37   &-21.70  &E    &Q  &   \\
\objectname[B2 1425+26]{B2 1425+267}      &0.366   &15.88  &17.47  &-25.49   &-23.45  &E    &L  &   \\
\objectname[MS 1426.5+0130]{MS 1426.5+0130}   &0.086   &14.30  &14.49  &-23.87   &-23.17  &S    &Q  &   \\
\objectname[PG 1444+407]{PG 1444+407}      &0.267   &15.80  &17.37  &-25.14   &-23.81  &S    &Q  &B  \\
\objectname[B2 1512+37]{B2 1512+37}       &0.371   &16.04  &17.31  &-25.38   &-23.66  &E    &L  &   \\
\objectname[MS 1519.8-0633]{MS 1519.8-0633}   &0.083   &16.01  &15.07  &-22.32   &-22.75  &S    &Q  &   \\
\objectname[3C 323.1]{3C 323.1}           &0.264   &16.07  &18.01  &-24.94   &-23.33  &E    &L  &B  \\
\objectname[MRC 1548+114]{MC 1548+114A}     &0.436   &18.27  &19.92  &-23.66   &-22.20  &SI   &L  &   \\
\objectname[MC2 1635+119]{MC 1635+119}      &0.146   &18.12  &16.73  &-21.38   &-22.62  &E    &Q  &   \\
\objectname[3C 351]{3C 351}             &0.372   &15.50  &16.97  &-25.96   &-24.59  &S    &L  &   \\
\objectname[PKS 2135-147]{PKS 2135-147}     &0.200   &16.21  &16.91  &-23.96   &-23.23  &E    &L  &   \\
\objectname[OX 169]{OX 169}             &0.211   &15.89  &17.28  &-24.59   &-23.18  &EI   &L  &   \\
\objectname[MS 2159.5-5713]{MS 2159.5-5713}   &0.083   &17.14  &15.01  &-20.91   &-22.52  &S    &?  &   \\
\objectname[HB89 2201+315]{Q 2201+315}       &0.295   &15.46  &16.75  &-25.78   &-24.50  &E    &L  &   \\
\objectname[LBQS 2214-1903]{LBQS 2214-1903}   &0.396   &18.81  &19.27  &-22.81   &-22.60  &S    &Q  &A  \\
\objectname[HB89 2215-037]{Q 2215-037}       &0.242   &18.69  &17.38  &-22.06   &-23.29  &E    &Q  &   \\
\objectname[PKS 2247+14]{PKS 2247+14}      &0.237   &16.65  &17.22  &-23.90   &-23.32  &E    &L  &   \\
\objectname[HB89 2344+184]{Q 2344+184}       &0.138   &20.22  &16.68  &-19.16   &-22.60  &S    &Q  &   \\
\objectname[PKS 2349-014]{PKS 2349-014}     &0.174   &15.97  &15.63  &-23.82   &-24.07  &IE   &L  &   \\

\enddata

\tablecomments {Col. (3), apparent nuclear magnitude in filter.
Col. (4), apparent host magnitude in filter.  Col. (5), absolute $V$
nuclear magnitude.  Col. (6), absolute $V$ host magnitude.  Col. (7),
host morphology: a) E=elliptical; b) S=spiral; c) EI=elliptical
undergoing strong interaction; d) SI=spiral undergoing strong
interaction; e) ED=elliptical with possible inner disk; f)
IE=irregular or interacting that is best fit with an elliptical model;
g) IS=irregular or interacting that is best fit with a spiral model.
Col. (8), radio-loudness: Q = radio-quiet; L = radio-loud; ? =
radio-loudness not available.  Col. (9), original proposal selections:
A = chosen from LBQS catalog; B = chosen from an optically-selected
catalog; C = chosen for lack of extended host in ground-based images.}

\tablenotetext {a} {Classified here as spiral because it has a small central bulge on top of 
a larger component with apparently tidal arms.  However, the tidal structure shows a 
de Vaucouleurs profile, and others (e. g., McLure et al.~1999) classify it as elliptical.}

\label{tab:fullsample}
\end{deluxetable}
%%%%%%%%%%%%%%%%%

\begin{deluxetable}{lccccccccccc}
\setlength{\tabcolsep}{0.02in}
\tabletypesize {\tiny}
\tablewidth{0pt}
\tablecaption{PCA Subsample}

\tablehead{
\colhead{$\begin{array}[t]{c} \mbox{Name}   \\                \\ (1) \end{array}$}&
\colhead{$\begin{array}[t]{c} \log r_{1/2}      \\  \\ (2) \end{array}$}&
\colhead{$\begin{array}[t]{c} \sigma_{\log r_{1/2}}  \\    \\ (3) \end{array}$}&
\colhead{$\begin{array}[t]{c} \mu_e  \\      \\ (4) \end{array}$}&
\colhead{$\begin{array}[t]{c} \sigma_{\mu_e} \\        \\ (5) \end{array}$}&
\colhead{$\begin{array}[t]{c} M_V \\ \mathrm{(bulge)}             \\ (6) \end{array}$}&
\colhead{$\begin{array}[t]{c} \sigma_{M_V} \\  \mathrm{(bulge)}   \\ (7) \end{array}$}&
\colhead{$\begin{array}[t]{c} M_V \\  \mathrm{(nuc)}              \\ (8) \end{array}$}&
\colhead{$\begin{array}[t]{c} \sigma_{M_V} \\ \mathrm{(nuc)}      \\ (9) \end{array}$}&
\colhead{$\begin{array}[t]{c} \log L_X    \\           \\ (10) \end{array}$}&
\colhead{$\begin{array}[t]{c} \sigma_{\log L_X}    \\    \\ (11) \end{array}$}&
\colhead{$\begin{array}[t]{c} \mbox{X-ray Ref.}    \\  \\ (12) \end{array}$}
}

%                      lg(r)         +-lg(r)    mu_e     +-mu     mv_bulge  +-bulge mv_nuc    +-nuc    log L_x   +-log L_x
\startdata
LBQS 0020+0018          &   0.471 &    0.024     &19.52   &  0.13  & -22.77  & 0.06  & -22.45  & 0.15   & 44.36   &  0.13 &1  \\
PHL 909                 &   0.919 &    0.005     &20.66   &  0.07  & -23.29  & 0.06  & -24.07  & 0.14   & 44.60   &  0.05 &1  \\
UM 301                  &   0.643 &    0.045     &20.69   &  0.24  & -22.44  & 0.08  & -23.92  & 0.15   & 44.66   &  0.08 &1  \\
3C 47                   &   0.731 &    0.023     &20.12   &  0.14  & -23.37  & 0.08  & -24.07  & 0.15   & 44.99   &  0.02 &2   \\
3C 48                   &   1.290 &    0.004     &21.06   &  0.10  & -24.83  & 0.10  & -25.72  & 0.14   & 45.34   &  0.01 &2   \\
PHL 1093                &   1.164 &    0.006     &22.01   &  0.07  & -23.54  & 0.06  & -23.47  & 0.15   & 44.45   &  0.01 &2   \\
PKS 0202-76             &   0.572 &    0.013     &19.53   &  0.12  & -23.12  & 0.10  & -24.98  & 0.14   & 44.93   &  0.09 &2   \\
Q 0244+194              &   0.889 &    0.012     &21.83   &  0.09  & -22.47  & 0.06  & -23.29  & 0.15   & 44.49   &  0.03 &1  \\
PKS 0312-77             &   1.313 &    0.003     &22.05   &  0.08  & -23.78  & 0.08  & -24.40  & 0.14   & 44.66   &  0.07 &2   \\
3C 93                   &   0.441 &    0.025     &18.28   &  0.15  & -23.73  & 0.08  & -23.51  & 0.17   & 44.73   &  0.04 &3  \\
IR 0450-2958    &   0.896 &    0.010     &21.83   &  0.15  & -22.51  & 0.14  & -25.41  & 0.14   & 44.70   &  0.02 &2   \\
PKS 0736+01             &   1.074 &    0.005     &21.70   &  0.06  & -23.58  & 0.06  & -24.03  & 0.14   & 44.22   &  0.03 &2   \\
MS 07546+3928           &   0.568 &    0.007     &19.25   &  0.10  & -23.47  & 0.10  & -24.22  & 0.15   & 43.58   &  0.15 &2   \\
IR 0759+6508    &   0.961 &    0.004     &21.29   &  0.14  & -23.34  & 0.13  & -23.57  & 0.14   & 42.28   &  0.15 &2   \\
MS 0801.9+2129  &   0.568 &    0.012     &20.49   &  0.17  & -22.13  & 0.16  & -22.91  & 0.14   & 43.40   &  0.22 &1  \\
3C 206                  &   1.484 &    0.020     &23.69   &  0.13  & -23.09  & 0.08  & -24.03  & 0.15   & 44.93   &  0.04 &2   \\
3C 215                  &   0.891 &    0.047     &20.94   &  0.25  & -23.08  & 0.10  & -24.08  & 0.14   & 44.57   &  0.03 &2   \\
PG 0923+201             &   1.090 &    0.010     &22.35   &  0.10  & -23.00  & 0.08  & -24.73  & 0.14   & 43.88   &  0.14 &1  \\
MS 0944.1+1333          &   0.789 &    0.009     &21.19   &  0.11  & -22.52  & 0.10  & -24.16  & 0.14   & 44.17   &  0.07 &1  \\
PG 1012+008     &   0.939 &    0.040     &21.61   &  0.31  & -22.64  & 0.24  & -23.75  & 0.15   & 44.04   &  0.13 &1  \\
HE 1029-1401            &   0.915 &    0.004     &21.59   &  0.08  & -22.74  & 0.07  & -24.79  & 0.14   & 44.83   &  0.01 &1  \\
PG 1202+281             &   0.656 &    0.006     &20.53   &  0.10  & -22.62  & 0.09  & -23.03  & 0.15   & 44.27   &  $<$0.01 &1  \\
PG 1216+069             &   1.080 &    0.035     &22.69   &  0.25  & -22.55  & 0.17  & -25.76  & 0.14   & 44.98   &  0.01 &2   \\
LBQS 1222+1235          &   1.167 &    0.018     &21.50   &  0.11  & -23.82  & 0.06  & -24.04  & 0.14   & 44.58   &  0.02 &1  \\
3C 273                  &   1.126 &    0.007     &21.06   &  0.10  & -24.24  & 0.10  & -27.19  & 0.14   & 45.52   &  $<$0.01 &2   \\
PG 1229+204     &   0.833 &    0.009     &21.52   &  0.15  & -22.45  & 0.14  & -22.27  & 0.14   & 44.07   &  0.01 &4  \\
PG 1302-102             &   1.129 &    0.008     &21.01   &  0.10  & -24.14  & 0.09  & -25.93  & 0.14   & 44.97   &  0.01 &2   \\
PG 1307+085             &   0.820 &    0.010     &21.57   &  0.11  & -22.42  & 0.10  & -24.33  & 0.15   & 44.41   &  0.01 &2   \\
MS 1416.3-1257          &   0.617 &    0.016     &21.24   &  0.13  & -21.70  & 0.10  & -23.37  & 0.14   & 44.43   &  0.01 &2   \\
B2 1425+267             &   1.322 &    0.006     &22.88   &  0.10  & -23.45  & 0.10  & -25.49  & 0.14   & 44.01   &  0.03 &2   \\
PG 1444+407     &   0.129 &    0.046     &17.04   &  0.35  & -22.96  & 0.26  & -25.14  & 0.15   & 44.56   &  0.02 &2   \\
MS 1519.8-0633  &   0.574 &    0.039     &21.01   &  0.31  & -21.42  & 0.24  & -22.32  & 0.15   & 43.90   &  0.11 &1  \\
3C 323.1                &   1.040 &    0.010     &21.61   &  0.09  & -23.33  & 0.08  & -24.94  & 0.14   & 44.93   &  0.01 &2   \\
MC 1635+119             &   0.846 &    0.004     &21.17   &  0.06  & -22.62  & 0.06  & -21.38  & 0.15   & 43.43   &  0.12 &2   \\
3C 351          &   1.016 &    0.012     &20.86   &  0.16  & -24.01  & 0.15  & -25.96  & 0.14   & 44.39   &  0.52 &2   \\
PKS 2135-147            &   1.158 &    0.007     &22.46   &  0.07  & -23.23  & 0.06  & -23.96  & 0.15   & 44.95   &  0.02 &2   \\
OX 169         &   0.885 &    0.003     &20.48   &  0.13  & -23.21  & 0.12  & -24.59  & 0.14   & 44.26   &  0.03 &2   \\
Q 2201+315              &   1.058 &    0.006     &20.30   &  0.08  & -24.50  & 0.08  & -25.78  & 0.14   & 45.16   &  0.04 &2   \\
Q 2215-037              &   1.006 &    0.004     &21.48   &  0.08  & -23.29  & 0.08  & -22.06  & 0.15   & 44.42   &  0.13 &1  \\
PKS 2247+14             &   1.231 &    0.004     &22.20   &  0.07  & -23.32  & 0.06  & -23.90  & 0.14   & 43.11   &  0.09 &2   \\
Q 2344+184     &   0.646 &    0.013     &20.91   &  0.15  & -22.11  & 0.14  & -19.16  & 0.29   & 43.5     &  0.1 &5  \\
PKS 2349-014            &   1.443 &    0.003     &22.89   &  0.06  & -24.07  & 0.06  & -23.82  & 0.15   & 44.71   &  0.03 &2   \\

\enddata

\tablecomments{
For elliptical hosts, we take the entire host as the ``bulge.''  For other host types, the bulge is the spheroidal component.  
Col. (2), half-light radius of host bulge (kpc).  Col. (4), effective
surface magnitude of host bulge (mag arcsec$^{-2}$).  Col. (6), absolute $V$ magnitude of
host bulge.  Col. (10), 0.5 keV x-ray luminosity of nucleus (erg s$^{-1}$).
}
\tablerefs{
X-ray literature; (1) Yuan et al.~1998; (2) Brinkmann et al.~1997; 
(3) Wilkes et al.~1994; (4) Grupe et al.~2001; (5) Margon et al.~1985.}
\label{tab:pcasample}
\end{deluxetable}

%-------------------------------------

\normalsize
\begin{deluxetable}{lllrrlrrrr}
\tabletypesize {\tiny}
\tablewidth{0pt}
\tablecaption{Correlations and Linear Fits}

\tablehead{
\colhead{Dependent}&
\colhead{Independent}&
\colhead{Subsample}&
\colhead{$N_{\mathrm{obj}}$}&
\colhead{$\rho$}&
\colhead{Probability}&
\colhead{Slope}&
\colhead{$\sigma_{\mathrm{slope}}$}&
\colhead{Intercept}&
\colhead{$\sigma_{\mathrm{intercept}}$}
}

\startdata

$M_V\mathrm{(host)}$  & $M_V\mathrm{(nuc)}$  & All & 56 &    0.350 &    0.01  &  0.30  & 0.09  & -15.9  & 2.2  \\
                      &                      & LE  & 22 &    0.434 &    0.04  &  0.29  & 0.08  & -16.6  & 1.9  \\
                      &                      & QE  & 16 &    0.394 &    0.13  &  0.26  & 0.17  & -16.7  & 4.2  \\
                      &                      & LS  &  4 &    0.800 &    0.20  &  0.77  & 0.14  &  -4.5  & 3.5  \\
                      &                      & QS  & 14 &    0.002 &    0.99  & -0.03  & 0.16  & -23.9  & 4.0  \\
                      &                      & L   & 26 &    0.578 &    $<$0.01  &  0.39  & 0.09  & -14.1  & 2.2  \\
                      &                      & Q   & 30 &    0.168 &    0.37  &  0.14  & 0.14  & -19.6  & 3.4  \\
                      &                      & E   & 38 &    0.379 &     0.02  &  0.32  & 0.10  & -15.4  & 2.5  \\
                      &                      & S   & 18 &    0.315 &    0.20  &  0.25  & 0.17  & -17.1  & 4.3  \\
\hline
$M_V\mathrm{(bulge)}$ & $\log \left( \mathcal{M}_{\mathrm{BH}}/ \mathcal{M}_{\sun} \right)$ & All & 20 &  -0.263  &   0.26  & -1.43  &  1.18    & -10.5 &   10.5 \\
                      &                      & LE  & 10 &   0.103  &   0.78  &  -0.57 &  0.92    & -18.6 &    8.2 \\
                      &                      & QE  &  7 &  -0.847  &   0.02 & -1.86  &  1.41    &  -6.1 &   12.5 \\
                      &                      & LS  &  1 &   ...    &   ...   & ...    &  ...     &   ... &   ...  \\
                      &                      & QS  &  2 &   ...    &   ...   & 0.16   &  $<$0.01 & -24.1 &    0.1 \\
                      &                      & L   & 11 &   0.059  &   0.86  & -0.47  &  0.73    & -19.4 &    6.5 \\
                      &                      & Q   &  9 &  -0.594  &   0.09 & -1.31  &  1.12    & -11.2 &    9.8 \\
                      &                      & E   & 17 &  -0.217  &   0.40  & -1.64  &  1.91    &  -8.6 &   17.2 \\
                      &                      & S   &  3 &  -0.500  &   0.67  & 0.19   &  0.14    & -24.7 &    1.1 \\
\hline
 $M_V\mathrm{(nuc)}$  & $\log \left( \mathcal{M}_{\mathrm{BH}} / \mathcal{M}_{\sun} \right)$ & All & 26 &  -0.378 &   0.06 & -1.98 &  1.16 &   -6.9 &  10.2  \\
                      &                      & LE  & 10 &  -0.164 &    0.65 & -2.75 &  4.09 &    0.1 &  37.0  \\
                      &                      & QE  &  7 &  -0.536 &    0.22 & -5.16 &  4.98 &   22.4 &  44.8  \\
                      &                      & LS  &  1 &     ... &     ... &  ...  &  ...  &    ... &  ...   \\
                      &                      & QS  &  8 &  -0.643 &    0.09 &  3.05 &  4.03 &  -50.1 &  33.9  \\
                      &                      & L   & 11 &  -0.196 &    0.56 & -2.27 &  2.23 &   -4.4 &  20.0  \\
                      &                      & Q   & 15 &  -0.318 &    0.25 & -1.72 &  1.64 &   -9.2 &  14.3  \\
                      &                      & E   & 17 &  -0.472 &   0.06 & -4.33 &  4.31 &   14.7 &  38.9  \\
                      &                      & S   &  9 &  -0.667 &   0.05 &  2.09 &  2.19 &  -42.0 &  18.4  \\
\enddata

\tablecomments {For elliptical hosts, we take the entire host as the ``bulge.''  For other host types, the bulge is the spheroidal component.  
These correlations include only those objects for which $M_V\mathrm{(nuc)} \leq -23.0$.  
The number of objects in each sample varies from one test to another, according to the available data. 
Col. (4), number of objects in subsample.  
Col.~(5), Spearman rank correlation coefficient.  
Col.~(6), the significance, listed as the probability that
$\left| \rho \right|$ could be this large or larger by chance alone.
Fits are of the form $\mathrm{Dependent} = \mathrm{slope} \times \mathrm {Independent} + \mathrm{intercept}$.}
\label{tab:spear-fits}
\end{deluxetable}

%-------------------------------------

%-------------------------------------

\begin{deluxetable}{lcccc}
\tabletypesize{\footnotesize}
\tablewidth{0pt}

\tablecaption{Black Hole Masses, Eddington Fractions, and Bulge Magnitudes}
\tablehead{
\colhead{Name}&
\colhead{$\log \mathcal{M}_{\mathrm{BH}} / \mathcal{M}_{\sun}$}&
\colhead{$\log L_{\mathrm{bol}} / L_{\mathrm{Edd}}$}&
\colhead{$M_V\mathrm{(bulge)}$}&
\colhead{$\sigma_{M_V\mathrm{(bulge)}}$}
}
\startdata

PG 0052$+$251    &8.8    &-0.8     &  ...    & ...   \\
PHL 909        &9.4    &-1.4   & -23.29  & 0.06  \\
PHL 1093         &9.1    &-1.3   & -23.54  & 0.06  \\

MRK 1014         &8.2    &-0.5     &  ...    & ...   \\
NAB 0205$+$02    &8.3    &-0.2     &  ...    & ...   \\
Q 0244$+$194     &8.5    &-0.8     & -22.47  & 0.06  \\

PKS 0736$+$01    &8.5    &-0.5     & -23.58  & 0.06  \\
PG 0923$+$201    &9.4    &-1.1   & -23.00  & 0.08  \\
PG 0953$+$414    &8.9    &-0.4     &  ...    & ...   \\

PKS 1004$+$13    &9.6    &-0.9     & -24.06  & 0.12  \\
PG 1012$+$008    &8.3    &-0.4     & -22.64  & 0.24  \\
HE 1029$-$1401   &9.6    &-1.2   & -22.74  & 0.07  \\

PG 1116$+$215    &8.7    &-0.2     &  ...    & ...   \\
PG 1202$+$281    &8.8    &-1.2   & -22.62  & 0.09  \\
3C 273           &9.1    &0.2       & -24.24  & 0.10  \\

PG 1302$-$102    &8.8     &0.0    & -24.14  & 0.09  \\
PG 1307$+$085    &8.3    &-0.2     & -22.42  & 0.10  \\
PG 1309$+$355    &8.5    &-0.3     & -23.62  & 0.17  \\

PG 1402$+$261    &7.8    &0.3        &  ...    & ...   \\
PG 1444$+$407    &8.5    &-0.1      & -22.96  & 0.26  \\
3C 323.1         &9.4    &-1.0   & -23.33  & 0.08  \\

MC 1635$+$119    &8.6    &-1.6  & -22.62  & 0.06  \\
PKS 2135$-$147   &9.4    &-1.4   & -23.23  & 0.06  \\
OX 169           &9.2    &-1.0     & -23.21  & 0.12  \\

PKS 2247$+$14    &8.1    &-0.1      & -23.32  & 0.06  \\
PKS 2349$-$014   &9.3    &-1.3   & -24.07  & 0.06  \\
\enddata

\tablecomments{
For elliptical hosts, we describe the entire host as the ``bulge.''  For other host types, the bulge is the spheroidal component.  
A standard deviation of 0.4 dex is used for all of the black hole masses, 
adopted from Vestergaard~(2004).  Since the errors in $M_V\mathrm{(nuc)}$ are much smaller,
the propagated errors in $\log \left ( L_{\mathrm{bol}} / L_{\mathrm{Edd}} \right )$ all come out to be 0.4, 
as well.}
\label{tab:bh}
\end{deluxetable}

%-------------------------------------

%-------------------
\begin{deluxetable}{ccrr}
\tabletypesize {\footnotesize}
\tablewidth{0pt}
\tablecaption{PCA Results}
\tablehead{
\colhead{$\begin{array}[t]{c} \mbox{Type}\\ \end{array}$}&
\colhead{$\begin{array}[t]{c} \mbox{Eigenvector}\\ \end{array}$}&
\colhead{$\begin{array}[t]{c} \lambda \\ \mbox{(\%)} \end{array}$}&
\colhead{$\begin{array}[t]{c} \mbox{Cumulative \%} \\ \end{array}$}}
\startdata

%class(#)  eigenvec        lambda(%)    cum.(%)
Optical & \bvec{e_1} & 63.4 &  63.4 \\
 & \bvec{e_2} & 32.6 &  96.1 \\
     & \bvec{e_3} &  3.9 & 100.0 \\
\hline
X-Ray  & \bvec{e_1} & 61.7 &  61.7 \\
     & \bvec{e_2} & 33.5 &  95.2 \\
     & \bvec{e_3} &  4.8 & 100.0 \\
\enddata
\tablecomments{Col.~(3), eigenvalue as a percent of total variance. Col.~(4), cumulative
percentages.}

\label{tab:pca-overall-optandx-42}
\end{deluxetable}

%-------------------
% calculated from: 
% ~/research/idl_codes_diss/fp_corr.pro
% ~/research/pca/newwork/rotation/fp_errors.pro (absolute errors)
\begin{deluxetable}{rcr}
\tablewidth{0pt}
\tablecaption{Fundamental Plane RMS Errors and Correlations}
\tablehead{
\colhead{Variable}&
\colhead{RMS Error}&
\colhead{Correlation}
}
\startdata

\multicolumn{3}{c}{Optical} \\
$\log r_{1/2}$    & 0.139  & 0.883 \\
$\mu_e$  & 0.630  & 0.866 \\
$M_V\mbox{(nuc)}$ & 1.97    & 0.532 \\
\hline		      		   
\multicolumn{3}{c}{X-ray} \\         
$\log r_{1/2}$        & 0.156 & 0.855 \\
$\mu_e$      & 0.668 & 0.847 \\
$\log L_X$ & 1.36 & 0.367 \\

\enddata

\tablecomments{
Optical fundamental plane results are listed first, and
x-ray FP results follow.  Col.~(2), RMS differences between the
measured variable and its FP solution. Col.~(3), Pearson correlation
coefficient between the measured variable and its FP solution. 
}

\label{tab:fp-rms}
\end{deluxetable}

%%%%%%%%%%%%%%%%%%%%%%%%%%%%%%%%%%%%%%%%%%%%%%%%%%%%%%%%%%
%APPENDIX A TABLES
%%%%%%%%%%%%%%%%%%%%%%%%%%%%%%%%%%%%%%%%%%%%%%%%%%%%%%%%%%

%%%%%%%%%%%%%%%%%%%%%%%%%%%%%%%%%%%%%%%%%%%%%%%%%%%%%%%%%%%

\begin{deluxetable}{lccccc}
\tabletypesize {\tiny}
\tablewidth{0pt}
\tablecaption{Comparison with the Literature}
\tablehead{
\colhead{Name}&
\colhead{$\Delta m_\mathrm{nuc}$}&
\colhead{$\Delta m_\mathrm{host}$}&
\colhead{$\Delta \log r_\mathrm{e}$}&
\colhead{Morphology}&
\colhead{Reference}
}

\startdata
% 1                2            3            4           5         6
%                  m_nuc        m_host       log r                    
%name              we-they      we-they      we-they     Mph       Ref
LBQS 0020$+$0018   &-0.22       & 0.01       &...        &E        &4 \\
LBQS 0021$-$0301   &-0.12       &-0.77       &...        &E        &4 \\
PG 0043$+$039      &-1.26       & 0.13       &-0.547     &E        &2 \\
PG 0052$+$251      &...         &-0.37       & 0.20      &S\tablenotemark{a}        &1 \\
PHL 909            &...         &-0.31       &-0.018     &E        &1 \\
UM 301             &-0.15       &-0.53       &...        &E        &4 \\
3C 48              &...         &-1.02       & 0.26      &E        &6 \\
PHL 1093           &-0.09       &-0.03       & 0.045     &E        &7 \\
PKS 0202$-$76      &-1.23       &-1.28       & 0.04      &E        &2 \\
NAB 0205$+$02      &...         &-0.92       & 0.30      &S\tablenotemark{a}        &1 \\
Q 0244$+$194       & 0.00       & 0.04       &-0.06      &E        &7 \\
US 3498            &-0.20       &-0.03       & 0.065     &S\tablenotemark{a}        &7 \\
PKS 0312$-$77      &-1.77       &-1.03       & 0.068     &E        &2 \\
PKS 0736$+$01      & 0.10       &-0.20       &-0.0496    &E        &7 \\
PG 0923$+$201      &...         &-0.04       & 0.00      &E        &1 \\
PKS 1004$+$13      &...         & 0.10       & 0.15      &E        &1 \\
HE 1029$-$1401     &...         &-0.34       & 0.08      &E        &1 \\
PG 1202$+$281      &...         &-0.31       &-0.056     &E        &1 \\
LBQS 1209$+$1259   &-0.21       &-0.82       &...        &E        &4 \\
LBQS 1218$+$1734   &-0.03       &-0.62       &...        &E        &4 \\
LBQS 1222$+$1010   &-0.08       &-1.97       &...        &S        &4 \\
LBQS 1222$+$1235   &-0.18       &-0.60       &...        &E        &4 \\
3C 273             &...         &-0.35       & 0.0       &E        &1 \\
PG 1229$+$204      &-0.13       & 0.44       &...        &S\tablenotemark{b}   &5 \\
LBQS 1240$+$1754   &-0.15       &-0.19       &...        &E        &4 \\
LBQS 1243$+$1701   &-0.26       &-2.40       &...        &E        &4 \\
PG 1302$-$102      &...         &-0.85       & 0.36      &E        &1 \\
PG 1307$+$085      &...         &-0.33       & 0.18      &E        &1 \\
PG 1309$+$355      &...         &-0.19       & 0.0       &S\tablenotemark{c}        &1 \\
PG 1358$+$04       &-1.24       &-1.78       &...\tablenotemark{d}        &E        &2 \\
PG 1402$+$261      &...         &-0.97       & 0.04      &S\tablenotemark{a}        &1 \\
B2 1425$+$267      &...         &-0.13       & 0.11      &E        &6 \\
3C 323.1           &...         &-0.09       & 0.11      &E        &1 \\
MC 1635$+$119      & 0.02       &-0.07       & 0.04      &E        &7 \\
3C 351             &-1.00       &-2.23       & 0.43      &S\tablenotemark{e}        &2 \\
OX 169             &-0.11       & 0.08       & 0.32      &E\tablenotemark{e}        &7 \\
LBQS 2214$-$1903   &-0.17       &-0.48       &...        &S        &4 \\
Q 2215$-$037       & 0.09       &-0.82       & 0.28      &E        &3 \\
PKS 2247$+$14      &-0.25       & 0.02       & 0.201     &E        &7 \\
Q 2344$+$184       & 1.02       &-0.52       & 0.08      &S\tablenotemark{a}        &7 \\
PKS 2349$-$014     &-0.03       &-0.27       & 0.177     &E        &7 \\

\enddata
\label{table:complist2}

\tablecomments {
In each of columns (2), (3), and (4), the difference is calculated as our value - the literature value.  
Col. (2), difference in nuclear magnitude.
Col. (3), difference in host magnitude.
Col. (4), difference in log effective radius 
($\log r_{1/2}$ for ellipticals or $\log r_\mathrm{eff}$ for spirals).  
Col. (5), host morphology, quoted in its simplest form: E (elliptical) or S (spiral).
The number of significant figures in the table varies with the precision of the literature values.  
Ellipses (...) are shown if no value is given.  
Values quoted from Bahcall et al.~(1997) and Kirhakos et al.~(1999) are their 2D model results.  
}
\tablerefs{
Literature references; 
  (1) Bahcall et al.~(1997); 
  (2) Boyce et al.~(1998); 
  (3) Disney et al.~(1995); 
  (4) Hooper et al.~(1997); 
  (5) Hutchings et al.~(1994a); 
  (6) Kirhakos et al.~(1999); 
  (7) McLure et al.~(1999).  
}
\tablenotetext {a} {Literature size assumes a disk only, 
while we account for both a bulge (or bar) and a disk.  
For PG~0052+251, NAB~0205+02, and PG~1402+261, we mask the central bulge or bar and fit only the disk, 
whereas Bahcall et al.~(1997) fit one exponential profile to the entire host.  
For Q~2344+184 and US~3498, we fit both bulge and disk separately and use our disk size here; 
McLure et al.~(1999) fit one exponential profile to the entire host.  
}
\tablenotetext {b} {Hutchings et al.~(1994a) do not list a morphology.}
\tablenotetext {c} {Size is $r_{1/2}$.  
A de Vaucouleurs profile fits best, but we classify it as a spiral based on appearance.  
We compare with the de Vaucouleurs model of Bahcall et al.~(1997).}
\tablenotetext {d} {We find the radial profile of PG~1358+04 to be a ``broken'' de Vaucouleurs 
profile with two different effective radii, so we do not compare $\log r_{1/2}$ here.}
\tablenotetext {e} {Literature size assumes a bulge model for the entire host.  
We use our bulge model size here but mask other features. 
For 3C~351, the bulge is surrounded by a ring feature that we mask from the fit but 
Boyce et al.~(1998) include.  
For OX~169, bulge is crossed by a large, arm-like feature we mask from the fit; 
we do not know exactly which areas are fitted by McLure et al.~(1999).
}

\end{deluxetable}

%%%%%%%%%%%%%%%%%%%%%%%%%%%%%%%%%%%%%%%%%%%%%%%%%%%%%%%%%%%%%%%%%%%%%%%%%%%%

%%%%%%%%%%%%%%%%%%%%%%%%%%%%%%%%%%%%%%%%%%%%%%%%%%%%%%%%%%
%APPENDIX B TABLES
%%%%%%%%%%%%%%%%%%%%%%%%%%%%%%%%%%%%%%%%%%%%%%%%%%%%%%%%%%

%-------------------------------------

% New results from "Anastasia", ~/pca/newwork/rotation/sample_42_paper2/{comb-fp-o.out, etc.}

\begin{deluxetable}{ccrrcrr}
\tabletypesize {\footnotesize}
\tablewidth{0pt}
\tablecaption{PCA Results by Subsample}
\tablehead{
\colhead{$\begin{array}[t]{c} \mbox{Sample (\#)}\\ \end{array}$}&
\colhead{$\begin{array}[t]{c} \mbox{Eigenvector}\\ \end{array}$}&
\multicolumn{2}{c}{Optical}&
\colhead{}&
\multicolumn{2}{c}{X-ray} \\ [.2ex] \cline{3-4} \cline{6-7}
\colhead{}&
\colhead{}&
\colhead{$\lambda$ (\%)}&
\colhead{Cumulative \%}&
\colhead{}&
\colhead{$\lambda$ (\%)}&
\colhead{Cumulative \%}
}
\startdata

%class(#)  eigenvec        lambda(%)    cum.(%)

LE (19)  & \bvec{e_1} & 63.6 &  63.6 &  & 66.1 &  66.1 \\
 & \bvec{e_2} & 34.3 &  97.9 &  & 31.4 &  97.4 \\
     & \bvec{e_3} &  2.1 & 100.0 &  &  2.6 & 100.0 \\
\hline		  
QE (14)  & \bvec{e_1} & 69.4 &  69.4 &  & 67.1 &  67.1 \\
 & \bvec{e_2} & 25.7 &  95.0 &  & 28.3 &  95.4 \\
     & \bvec{e_3} &  5.0 & 100.0 &  &  4.6 & 100.0 \\
\hline		  
QS (8)   & \bvec{e_1} & 65.8 &  65.8 &  & 69.5 &  69.5 \\
 & \bvec{e_2} & 32.4 &  98.1 &   & 28.0 &  97.4 \\
     & \bvec{e_3} &  1.9 & 100.0 &  &  2.6 & 100.0 \\
\hline		  
L (20)   & \bvec{e_1} & 63.7 &  63.7 &  & 65.8 &  65.8 \\
 & \bvec{e_2} & 34.2 &  97.9 &  & 31.6 &  97.4 \\
     & \bvec{e_3} &  2.1 & 100.0 &  &  2.6 & 100.0 \\
\hline		  
Q (22)   & \bvec{e_1} & 63.6 &  63.6 &  & 63.1 &  63.1 \\
 & \bvec{e_2} & 33.0 &  96.6 &  & 33.3 &  96.3 \\
     & \bvec{e_3} &  3.4 & 100.0 &  &  3.7 & 100.0 \\
\hline		  
E (33)   & \bvec{e_1} & 64.5 &  64.5 &  & 61.1 &  61.1 \\
 & \bvec{e_2} & 30.8 &  95.3 &  & 33.8 &  94.8 \\
     & \bvec{e_3} &  4.7 & 100.0 &  &  5.2 & 100.0 \\
\hline		  
S (9)    & \bvec{e_1} & 62.2 &  62.2 &  & 65.4 &  65.4 \\
 & \bvec{e_2} & 35.4 &  97.7 &  & 30.0 &  95.5 \\
     & \bvec{e_3} &  2.3 & 100.0 &  &  4.5 & 100.0 \\
\enddata

\tablecomments{Col. (1), subsample and number of QSOs.  Cols. (3) \& (5),
eigenvalue as a percent of total variance.  Cols. (4) \& (6), cumulative
percentages.}

\label{tab:pca-comprehensive-42}
\end{deluxetable}

%-------------------------------------

%-------------------------------------

%from: ofp_params.table
%and   xfp_params.table
\begin{deluxetable}{ccrrr}
\tablewidth{0pt}
\tablecaption{Coefficients for FP Forms}
\tablehead{
\colhead{Sample}&
\colhead{$R_{\mathrm{coeff}}$}&
\colhead{$M_{\mathrm{coeff}}$}&
\colhead{$C$}
}
\startdata
%             r            m           const
\multicolumn{4}{c}{Optical} \\
All &   -14.2  &    3.14  &   -77.5  \\
LE &   -14.7  &    3.14  &   -75.8  \\
QE &    108  &  -22.4   &   361    \\
QS & -29.5    &  5.55     & -118    \\
L &   -15.3  &    3.29  &   -78.3  \\
Q &   -44.9  &   8.53   &  -168    \\
E &   -14.1  &    3.13  &   -77.1  \\
S &   -19.1  &    3.79  &   -88.1  \\
\hline		      		   
\multicolumn{4}{c}{X-ray} \\
All &    8.74  &  -2.03  &  79.3  \\
LE &   10.8   &  -2.39  &  84.1  \\
QE &  -11.8   &   2.55  &   0.0918 \\
QS &  -33.5   &   5.89  & -54.9  \\
L  &   11.4   &  -2.51  &  85.9  \\
Q  &    -59.3 &   11.5  & -151  \\
E  &   8.14   &  -1.93  &  77.6  \\
S &   42.1  &   -8.36  & 186  \\
\enddata

\tablecomments{
Coefficients and constants for the fundamental plane equations:  
in the optical case, $M_V\mathrm{(nuc)}=R_{\mathrm{coeff}} \log r_{1/2}+M_{\mathrm{coeff}} \mu_e+C$, and in the x-ray case, $\log L_X=R_{\mathrm{coeff}} \log r_{1/2}+ M_{\mathrm{coeff}} \mu_e+C$.
}

\label{tab:fp-physicalforms}
\end{deluxetable}

%-------------------------------------

%-------------------------------------

%NB: output from fp_angles2.sm
\begin{deluxetable}{lcrr}
\setlength{\tabcolsep}{0.2in}
%\tabletypesize {\tiny}
\tablewidth{0pt}
\tablecaption{QSO FP Gradients}
\tablehead{
\colhead{Sample}&
\colhead{$\begin{array}[t]{c} \mbox{Gradient magnitude } \left| \left| \nabla F \right| \right|  \\  \end{array}$}&
\colhead{$\begin{array}[t]{c} \mbox{Azimuth } \alpha \\ \mbox{(deg)} \end{array}$}
}
\startdata
%sample(no)   gradient  azimuth
\multicolumn{3}{c}{Optical} \\
\bfseries All &  \bfseries 6.49  &  \bfseries 151  \\
Q   &  19.9   &  155  \\
QS  &  13.1   &  155  \\
S   &  8.52   &  154  \\
L   &  6.95   &  152  \\
LE  &  6.65   &  152  \\
E   &  6.45   &  151  \\
QE  &  48.7   &  -27  \\
\hline		      		   
\multicolumn{3}{c}{X-ray} \\
\bfseries All & \bfseries 10.1   & \bfseries -30 \\
S   & 47.0   &   -26 \\
L   & 13.0   &   -29 \\
LE  & 12.4   &   -29 \\
E   & 9.46   &   -31 \\
QE  & 13.4   &   152 \\
QS  & 36.6   &   156 \\
Q   & 65.9   &   154 \\
\enddata

\tablecomments{
Full sample results are given in boldface.  
The subsample QSO FPs are grouped by azimuth and ranked by magnitude.  
The gradients are unitless, while the azimuthal directions are measured in degrees 
counterclockwise from the $+ \bvec{i}$ axis.  
}

\label{tab:fp-gradients}
\end{deluxetable}

%%%%%%%%%%%%%


\clearpage

\begin{thebibliography}{}

\bibitem[Akritas \& Bershady(1996)]{Akritas96} Akritas, M. G., \&
Bershady, M. A. 1996, \apj, 470, 706

\bibitem[Bahcall et al.(1997)]{Bahcall97} Bahcall, J.  N., Kirhakos,
S., \& Saxe, D.  H. 1997, \apj, 479, 642

\bibitem[Bernardi et al.(2007)]{Bernardi07} Bernardi, M., Sheth, R. K., Tundo, E., \& Hyde, J. B. 2007, \apj, 660, 267

\bibitem[Blandford(1999)]{Blandford99} Blandford, R. 1999, in ASP
Conf. Ser. 182, Galaxy Dynamics, ed. D. R. Merritt, M. Valluri, \&
J. A. Sellwood (San Francisco: ASP), 87

\bibitem[Bondi(1952)]{Bondi52} Bondi, H. 1952, \mnras, 112, 195


\bibitem[Boyce et al.(1998)]{Boyce98} Boyce, P. J., Disney, M. J., Blades, J. C., Boksenberg, A., Crane, P., Deharveng, J. M., Macchetto, F. D., Mackay, C. D., \& Sparks, W. B. 1998, \mnras, 298, 121 


\bibitem[Boyce et al.(1993)]{Boyce93} Boyce, P. J., Phillipps, S., \& Davies, J. I. 1993, \aap, 280, 694 


\bibitem[Brinkmann et al.(1997)]{Brinkmann97} Brinkmann, W., Yuan, W.,
\& Siebert, J. 1997, \aap, 319, 413

\bibitem[Carter(1979)]{Carter79} Carter, B. 1979, in Active Galactic
Nuclei, ed. C. Hazard \& S. Mitton (Cambridge: Cambridge Univ. Press),
185

\bibitem[Cristiani \& Vio(1990)]{CV90} Cristiani, S. \& Vio, R. 1990,
\aap, 227, 385

\bibitem[Di Matteo et al.(2000)]{DiMatteo00} Di Matteo, T., Quataert,
E., Allen, S. W., Narayan, R., \& Fabian, A. C. 2000, \mnras, 311, 507

\bibitem[Disney et al.(1995)]{Disney95} Disney, M. J., Boyce, P. J., Blades, J. C., Boksenberg, A., Crane, P., Deharveng, J. M., Macchetto, F., Mackay, C. D., Sparks, W. B., \& Phillipps, S. 1995, Nature, 376, 150


\bibitem[Djorgovski \& Davis(1987)]{Djorgovski87} Djorgovski, S., \&
Davis, M. 1987, \apj, 313, 59

\bibitem[Dressler et al.(1987)]{Dressler87} Dressler, A., Lynden-Bell,
D., Burstein, D., Davies, R. L., Faber, S. M., Terlevich, R. J.,
Wegner, G. 1987, \apj, 313, 42

\bibitem[Dunlop et al.(2003)]{Dunlop03} Dunlop, J. S., McLure, R. J.,
Kukula, M. J., Baum, S. A., O'Dea, C. P., \& Hughes, D. H. 2003,
MNRAS, 340, 1095

\bibitem[Eggers et al.(2000)]{Eggers00} Eggers, D., Shaffer, D. B., \& Weistrop, D.
2000, \aj, 119, 460

\bibitem[Elvis et al.(1994)]{Elvis94} Elvis, M., et al. 1994, \apjs, 95, 1

\bibitem[Fabian(1999)]{Fabian99} Fabian, A. C. 1999, \mnras, 308, L39

\bibitem[Ferrarese \& Merritt(2000)]{Ferrarese00} Ferrarese, L., \&
Merritt, D. 2000, \apj, 539, L9

\bibitem[Ferrarese \& Ford(2005)]{Ferrarese05} Ferrarese, L., \&
Ford, H. 2005, Space Sci. Rev., 116, 523

\bibitem[Foltz et al.(1987)]{Foltz87} Foltz, C. B., Chaffee, F. H.,
Jr., Hewett, P. C., MacAlpine, G. M., Turnshek, D. A., Weymann, R. J.,
Anderson, S. F. 1987, \aj, 94, 1423

\bibitem[Fukugita et al.(1995)]{Fukugita95} Fukugita, M., Shimasaku,
K., \& Ichikawa, T. 1995, \pasp, 107, 945

\bibitem[Gebhardt et al.(2000)]{Gebhardt00} Gebhardt, K., et al. 2000, \apjl, 539, 13

\bibitem[Graham(2007)]{Graham07} Graham, A. W. 2007, \mnras, in press (arXiv:0705.0618)

\bibitem[Grupe et al.(2001)]{Grupe01} Grupe, D., 
Thomas, H.-C., \& Beuermann, K. 2001, \aap, 367, 470

\bibitem[Hamilton(2001)]{Hamilton01} Hamilton, T. S. 2001, PhD thesis,
Univ. of Pittsburgh

\bibitem[Hamilton et al.(2002)]{Hamilton02} Hamilton,
T. S., Casertano, S., \& Turnshek, D. A.  2002, \apj, 576, 61

\bibitem[H\"aring \& Rix(2004)]{Haring04} H\"aring, N., \& Rix, H. 2004, \apjl, 604, 89

\bibitem[Hooper et al.(1997)]{Hooper97} Hooper, E. J., Impey, C. D., \& Foltz, C. B.
1997, \apjl, 480, 95

\bibitem[Hutchings et al.(1984)]{Hutchings84}
Hutchings, J. B., Crampton, D., \& Campbell, B. 1984, \apj, 280, 41

\bibitem[Hutchings et al.(1994a)]{Hutchings94a}
Hutchings, J. B., Holtzman, J., Sparks, W. B., Morris, S. C., Hanisch, R. J., \& Mo, J.
1994a, \apjl, 429, 1

\bibitem[Hutchings et al.(1994b)]{Hutchings94b}
Hutchings, J. B., Morris, S. C., Gower, A. C., \& Lister, M. L. 
1994b, \pasp, 106, 642

\bibitem[J{\o}rgensen et al.(1996)]{Jorgensen96}
J{\o}rgensen, I., Franx, M., \& Kj{\ae}rgaard, P. 1996, \mnras, 280,
167

\bibitem[Kaspi et al.(2000)]{Kaspi00} Kaspi, S., Smith, P. S., Netzer,
H., Maoz, D., Jannuzi, B. T., \& Giveon, U. 2000, \apj, 533, 631

\bibitem[Kirhakos et al.(1999)]{Kirhakos99} Kirhakos, S., Bahcall, J. N., Schneider, D. P., \& Kristian, J.
1999, \apj, 520, 67

\bibitem[Kormendy(1977)]{Kormendy77} Kormendy, J. 1977, \apj, 218, 333

\bibitem[Kormendy(2001)]{Kormendy01} Kormendy, J. 2001, in ASP
Conf. Ser. 230, Galaxy Disks and Disk Galaxies, ed. J. G. Funes 
\& E. M. Corsini (San Francisco: ASP), 247

\bibitem[Kormendy \& Gebhardt(2001)]{KormendyGeb01} Kormendy, J., \& Gebhardt, K. 2001, in AIP Conf. Proc. 586, 20th Texas Symposium on Relativistic Astrophysics (Melville: AIP), 363

\bibitem[Krist \& Hook(2004)]{Krist04} Krist, J., \& Hook, R. 2004,
The Tiny Tim User's Guide, (Baltimore: STScI), 
http://www.stsci.edu/software/tinytim/tinytim.pdf

\bibitem[Laor(1998)]{Laor98} Laor, A. 1998, \apjl, 505, 83

\bibitem[Laor et al.(1997)]{Laor97} Laor, A., Fiore, F., Elvis, M.,
Wilkes, B., \& McDowell, J. 1997, \apj, 477, 93

\bibitem[Magorrian et al.(1998)]{Magorrian98} Magorrian, J., et al. 1998,
\apj, 115, 2285

\bibitem[Malkan(1984)]{Malkan84} Malkan, M. A. 1984, \apj, 287, 555

\bibitem[Marconi \& Hunt(2003)]{Marconi03} Marconi, A., \& Hunt, L. K. 2003, \apjl, 589, 21

\bibitem[Margon et al.(1985)]{Margon85} Margon, B., Downes, R. A.,
Chanan, G. A. 1985, \apjs, 59, 23

\bibitem[Mathews \& Ferland(1987)]{Mathews87} Mathews, W., \& Ferland,
G. 1987, \apj, 323, 456

\bibitem[McCray(1979)]{McCray79} McCray, R. 1979, in Active Galactic
Nuclei, ed. C. Hazard \& S. Mitton (Cambridge: Cambridge Univ. Press),
227

\bibitem[McLeod \& Rieke(1994a)]{McLeod94a} McLeod, K. K., \& Rieke,
G. H. 1994a, \apj, 420, 58

\bibitem[McLeod \& Rieke(1994b)]{McLeod94b} McLeod, K. K., \& Rieke,
G. H. 1994b, \apj, 431, 137

\bibitem[McLure \& Dunlop(2001)]{McLure01} McLure, R. J., \& Dunlop,
J. S. 2001, \mnras, 327, 199

\bibitem[McLure \& Dunlop(2002)]{McLure02} McLure, R. J., \& Dunlop,
J. S. 2002, \mnras, 331, 795

\bibitem[McLure et al.(1999)]{McLure99} McLure, R. J., Kukula, M. J., 
Dunlop, J. S., Baum, S. A., O'Dea, C. P., \& Hughes, D. H 
1999, \mnras, 308, 377


\bibitem[McLure et al.(2006)]{McLure06} McLure, R. J., Jarvis, M. J., Targett, T. A., Dunlop, J. S., \& Best, P. N. 2006, MNRAS, 368, 139

\bibitem[Merritt \& Ferrarese(2001)]{Merritt01} Merritt, D., \& 
Ferrarese, L. 2001, \apj, 547, 140

\bibitem[Mushotzky(1997)]{Mushotzky97} Mushotzky, R. F. 1997, in ASP
Conf. Ser. 128, Mass Ejection from Active Galactic Nuclei, ed. N. Arav,
I. Shlosman, \& R. J. Weedman (San Francisco: ASP), 141

\bibitem[O'Dowd et al.(2002)]{ODowd02} O'Dowd, M., Urry,
C. M., \& Scarpa, R. 2002, ApJ, 580, 96

\bibitem[Pence(1976)]{Pence76} Pence, W. 1976, \apj, 203, 39

\bibitem[Remy et al.(1997)]{Remy97} Remy, M., Surdej, J., Baggett, S.,
\& Wiggs, M. 1997, in 1997 HST Calibration Workshop, ed. S. Casertano
et al. (Baltimore: STScI), 374

\bibitem[Schlegel et al.(1998)]{Schlegel98} Schlegel, D. J., Finkbeiner, 
D. P., \& Davis, M. 1998, \apj, 500, 525

\bibitem[Schmidt \& Green(1983)]{Schmidt83} Schmidt, M., \& Green,
R. F. 1983, \apj, 269, 352

\bibitem[Scodeggio et al.(1998)]{Scodeggio98} Scodeggio, M., Gavazzi,
G., Belsole, E., Pierini, D., \& Boselli, A. 1998, \mnras, 301, 1001

\bibitem[Silk \& Rees(1998)]{Silk98} Silk, J., \& Rees, M. J. 1998,
\aap, 331, L1

\bibitem[Tremaine et al.(2002)]{Tremaine02} Tremaine, S., et al. 2002, \apj, 574, 740

\bibitem[van Albada et al.(1993)]{vanAlbada93} van
Albada, T. S., Bertin, G., \& Stiavelli, M. 1993, \mnras, 265, 627

\bibitem[Vestergaard(2004)]{Vestergaard04} Vestergaard, M. 2004, 
in ASP Conf. Ser. 311, AGN Physics with the Sloan Digital Sky Survey, 
ed. G. Richards \& P. Hall (San Francisco: ASP), 69

\bibitem[Voit(1997)]{Voit97} Voit, M., ed. 1997, HST Data Handbook,
Vol. I (Version 3.0; Baltimore: STScI)

\bibitem[Wilkes et al.(1994)]{Wilkes94} Wilkes, B. J., Tananbaum, H.,
Worrall, D. M., Avni, Y., Oey, M. S., \& Flanagan, J. 1994, \apjs, 92,
53

\bibitem[Woo \& Urry(2002)]{Woo02} Woo, J.-H., \& Urry, C. M. 2002,
\apj, 579, 530

\bibitem[Yuan et al.(1998)]{Yuan98} Yuan, W., Brinkmann, W., Siebert,
J., Voges, W. 1998, \aap, 330, 108

\bibitem[Zheng et al.(1997)]{Zheng97} Zheng, W., Kriss, G., Telfer,
R., Grimes, J., \& Davidsen, A. F. 1997, \apj, 475, 469







\end{thebibliography}
\end {document}